\documentclass[twocolumn]{aastex631}
\usepackage{color}
\usepackage{amsmath}

\newcommand{\ie}{i.e.,\ }
\newcommand{\eg}{e.g.,\ }
\newcommand{\etal}{et~al.\ }
\newcommand{\kms}{km~s$^{-1}$}
\newcommand{\magsec}{mag arcsec$^{-2}$}
\def\brackmueg{$\langle \mu_g \rangle_e$}
\def\Msun{$M_\odot$}
\def\ugi{$u^*\llap,\ g^\prime\llap,\ i^\prime$}
\def\galfit{{\tt galfit}}

\defcitealias{Mei07}{M07}
\defcitealias{Blakeslee09}{B09}

\begin{document}

\title{The Globular Cluster System of the Virgo Cluster Ultradiffuse Galaxy VCC~615}
\shorttitle{Globular Cluster System of VCC~615}
\shortauthors{Mihos \etal}

\author{J. Christopher Mihos}
\affiliation{Department of Astronomy, Case Western Reserve University, Cleveland OH 44106, USA}

\author{Patrick R. Durrell}
\affiliation{Department of Physics, Astronomy, Geophysics and Environmental Sciences, Youngstown State University, Youngstown, OH 44555 USA}

\author{Elisa Toloba}
\affiliation{Department of Physics and Astronomy, University of the Pacific, Stockton, CA 95211, USA}

\author{Eric W. Peng}
\affiliation{NOIRLab, 950 North Cherry Ave, Tucson, AZ, 85719, USA}

\author{Sungsoon Lim}
\affiliation{Department of Astronomy, Yonsei University, 50 Yonsei-ro Seodaemun-gu, Seoul, 03722, Republic of Korea}

\author{Patrick C\^ot\'e}
\affiliation{National Research Council of Canada, Herzberg Astronomy and Astrophysics Program, Victoria, BC V9E 2E7, Canada}

\author{Puragra Guhathakurta}
\affiliation{UCO/Lick Observatory, University of California Santa Cruz, Santa Cruz, CA 95064, USA}

\author{Laura Ferrarese}
\affiliation{National Research Council of Canada, Herzberg Astronomy and Astrophysics Program, Victoria, BC V9E 2E7, Canada}

\begin{abstract}

We use {\sl Hubble Space Telescope} imaging to study the globular
cluster system of the Virgo Cluster ultradiffuse galaxy (UDG) VCC~615.
We select globular cluster candidates through a combination of size and
color, while simultaneously rejecting contamination from background
galaxies that would be unresolved in ground-based imaging. Our sample of
globular cluster candidates is essentially complete down to a limiting
magnitude of F814W=24.0, $\approx$ 90\% down the globular cluster
luminosity function. We estimate a total globular cluster population for
VCC~615 of $N_{\rm GC}=25.1^{+6.5}_{-5.4}$, resulting in a
specific frequency of $S_N=55.5^{+14.5}_{-12.0}$, quite
high compared to normal galaxies of similar luminosity, but consistent
with the large specific frequencies found in some other UDGs. The
abundant cluster population suggests the galaxy is enshrouded
by a massive dark halo, consistent with previous
dynamical mass estimates using globular cluster kinematics. While
the peak of the globular cluster luminosity function appears slightly
brighter than expected (by $\approx$ 0.3--0.5 mag), this difference is
comparable to the 0.3 mag uncertainty in the measurement, and we see no
sign of an extremely luminous population of clusters similar to those
detected in the UDGs NGC1054-DF2 and -DF4. However, we do find a
relatively high fraction ($32^{+5}_{-4}$\%) of large clusters with
half-light radii $R_h>9$ pc. The galaxy's offset nucleus appears
photometrically distinct from the globular clusters, and is more
akin to ultracompact dwarfs (UCDs) in Virgo. Over time, VCC615's already diffuse stellar body
may be further stripped by cluster tides, leaving the nucleus intact to form
a new Virgo UCD.

\end{abstract}


\section{Introduction}

The nature of ultradiffuse galaxies (UDGs) continues to pose complicated
questions about the formation and evolution of galaxies. UDGs are large
and extremely diffuse galaxies first identified in dense galaxy clusters
\citep{sb84, binggeli87, vandokkum15, mihos15}, but are now known to
populate a variety of environments, including field and group
environments \citep{barbosa20,marleau21,zaritsky23}. These objects show
a wide range of properties, from the ``red and dead'' gas-poor UDGs
typically found in denser environments to those with bluer colors and
surprisingly high gas fractions \citep{cannon15,leisman17,mihos18}. This
diversity in the UDG population has led to a variety of proposed
formation scenarios. Some invoke processes that are intrinsic to the
galaxy upon formation, such as having a halo of high spin or low
concentration \citep{amorisco16,benavides23}, a truncated star formation
history \citep{peng16, vandokkum15, vandokkum16, janssens22}, or having
experienced overly-disruptive feedback from bursty star formation
\citep{dicintio17}. Other scenarios rely on subsequent evolutionary
processes such as tidal stripping and heating \citep{moore96, liao19,
carleton19} to transform otherwise normal galaxies into UDGs at later
times. These processes are not mutually exclusive, and both intrinsic
formation scenarios and subsequent evolution can in principle combine to
shape UDGs in a wide variety of ways. In fact, these processes may
provide an evolutionary link between UDGs and ultracompact dwarf
galaxies (UCDs) in dense clusters, if nucleated UDGs are completely
disrupted by cluster tides leaving only their compact nuclei behind
\citep[\eg][]{bekki03, pfeffer13, liu15, liu20, janssens19, wang23}.

The dark matter content of UDGs is of particular importance in
disentangling these different scenarios --- do UDGs have low mass
dwarf-like halos, or are they cocooned in massive dark halos
characteristic of more luminous galaxies? Dynamical mass estimates of
UDGs show a range of possibilities. Some UDGs have very high
mass-to-light ratios \citep{beasley16, toloba18, forbes21, toloba23},
indicating the presence of an overly-massive dark halo, while other UDGs
have much lower dynamical mass-to-light ratios \citep{vandokkum19,
chilingarian19, toloba23}, including some objects in which the stellar
mass dominates to total mass \citep{vandokkum18, danieli19, toloba23}.
Here again, the diversity of dynamical properties points to UDGs being a
heterogenous class of objects arising from multiple formation scenarios.

Into this picture, globular clusters provide important and complementary
information about the properties of UDGs. UDGs have a wide range of
globular cluster specific frequencies
\citep[$S_N=N_{GC}10^{0.4(M_V+15)}$;][]{hvdb81}; some have abundant
cluster systems \citep{peng16, vandokkum17, lim18, muller21, saif21,
danieli22, fielder23, fg24}, while others show sparser cluster
populations or sometimes no clusters at all \citep{amorisco18, prole19,
lim20,forbes20,marleau24}. In the normal galaxy population, multiple
studies have shown a correlation between the number of globular clusters
($N_{GC}$) or total {\it mass} of globular clusters ($M_{GC}$) in a
galaxy and its total dynamical mass \citep[\eg][]{blakeslee97b,peng08,
spitler09, georgiev10, harris13,
hudson14,harris17,burkert20,zaritsky22}. As shown by those studies, this
correlation holds over a wide range of galaxy mass, from dwarf galaxies
to massive ellipticals and cD galaxies. If UDGs also follow these
relationships, the globular cluster-rich UDGs are much more dark-matter
dominated than their globular cluster-poor counterparts, again arguing
for a variety of formation channels for UDGs. However, whether these
differences in cluster populations truly trace differences in dark
matter content remains unclear; in objects where both dynamical masses
and globular cluster counts are available, these two mass estimates are
sometimes discrepant \citep[\eg][]{vandokkum19}. Nonetheless, if the
connection between globular clusters and dark halo mass can be shown
universally to hold for UDGs, it provides an opportunity to estimate UDG
halo masses through imaging alone, without the need for
resource-intensive kinematic studies.

However, several factors complicate the study of globular clusters in
these diffuse galaxies, particularly in ground-based imaging which often
suffer from significant incompleteness and contamination effects. For
example, at the distance of Virgo \citep[d=16.5
Mpc][]{mei07,cantiello24}, ground based imaging only reaches the upper
half of the globular cluster luminosity function (GCLF), while also
being subject to significant contamination both from foreground Milky
Way stars and unresolved background galaxies. The large statistical
corrections for these effects, coupled with the low {\sl total} $N_{GC}$
values in UDGs (as opposed to their luminosity-normalized specific
frequency $S_N$) leads to significant uncertainties in deriving the
total globular cluster population of UDGs from the observed counts. For
example, in the well-studied UDG DF44, estimates of the total globular
cluster count have varied by factors of 4--5 between different studies
\citep[][see also \citealt{fg24}]{vandokkum16,vandokkum17,saif21}.
Furthermore, there is growing evidence that the cluster populations in
UDGs may show systematic differences from those of normal galaxies: some
UDGs have overly-luminous globular clusters \citep{vandokkum18, shen21a,
janssens22}, or globular cluster systems with abnormally low scatter in
color \citep{vandokkum22, janssens22}. Thus, the combination of
statistical uncertainty, contamination, and possible intrinsic
differences clouds the use of globular clusters for studying the
distances and masses of UDGs.

In this paper, we use deep {\sl Hubble Space Telescope} imaging to study
the globular cluster populations in the ultradiffuse galaxy VCC~615,
located in the outskirts of the Virgo Cluster. VCC~615 (detailed in
Table~\ref{props}) meets the structural definition of a UDG in the
analysis by \citet{lim20} who examined the photometric properties of
galaxies in the {\sl Next Generation Virgo Cluster Survey} galaxy sample
\citep{ferrarese20} and identified galaxies that were 2.5$\sigma$
outliers in a combination of size, luminosity, and surface brightness.
Several arguments make VCC~615 particularly well-suited for a study of
its globular cluster system. First, our previous work has pinpointed its
distance using the tip of the red giant branch (TRGB) method
\citep{mihos22}, putting it on the far side of Virgo at
$d=17.7^{+0.7}_{-0.5}$~Mpc, near the cluster virial radius. The low
$\approx$ 0.05 mag uncertainty in the distance modulus allows us to
derive accurate physical sizes and luminosities of it cluster
population. Furthermore, we have well-determined estimates of both the
galaxy's stellar mass
\citep[$M_*=7.3\pm1.1\times10^7$\Msun][]{roedigerIP} and dynamical mass
\citep[$M_{dyn}=2.3^{+2.8}_{-2.2}\times10^9$ \Msun, measured within the
effective radius of the globular cluster system, see ][]{toloba23} from
ground-based imaging and spectroscopy, respectively, allowing us to test
the connection between its cluster system and its halo mass. Finally,
using our {\sl Hubble} imaging we can identify globular clusters via a
combination of color and image concentration, while simultaneously
rejecting contamination from background galaxies which would be
unresolved in ground-based imaging. We use these data to construct a
very clean sample of globular clusters that probe the full range of the
GCLF, giving us the ability to study the cluster system of VCC~615 ---
both its intrinsic properties and its connection to its host galaxy ---
at a very high level of accuracy. Finally, we also use the data to study
the galaxy's compact nucleus, to evaluate scenarios that propose the
evolution of nucleated UDGs into UCDs, driven by tidal stripping in the
dense environment of the Virgo Cluster.

\begin{deluxetable}{ccc}
\tabletypesize{\small}
\tablewidth{0pt}
\tablecaption{VCC~615 \label{props}}
\tablehead{\colhead{Property} & \colhead{Value} & \colhead{Source}}
\startdata
Center (J2000) & (12:23:04.6, +12:00:53) & [1] \\
$m_{g'}$ & 17.25 & [1] \\
R$_{e,*}$ & 26.3\arcsec\ & [1] \\
\brackmueg & 26.3 \magsec & [1] \\
Distance & $17.7^{+0.7}_{-0.5}$ Mpc & [2] \\
$M_{g'}$ & $-$14.1 &  [1,2] \\
R$_{e,*}$ & 2.25 kpc & [1,2] \\
$v_{sys}$& $2089^{+16}_{-15}$ \kms\  & [3] \\
$M_*$ & $7.3\pm1.1 \times 10^7$ \Msun & [4] \\
$M_{dyn,1/2}$ & $2.3^{+2.8}_{-2.2} \times 10^9$ \Msun & [3] \\
\enddata
\tablerefs{
[1] \citet{lim20},
[2] \citet{mihos22},
[3] \citet{toloba23},
[4] \citet{roedigerIP}
}
\label{props}
\end{deluxetable}

\section{Observational Data and Image Reduction}

We imaged VCC~615 using the F475W and F814W filters of the Wide Field
Channel (WFC) of the Advanced Camera for Surveys (ACS) on the {\sl
Hubble Space Telescope} as part of program GO-15258 (Cycle 25). The goal
of the F814W imaging was to perform individual point source photometry
(using DOLPHOT) of faint red giant branch (RGB) stars to reach below the
TRGB to obtain a distance estimate to VCC~615 \citep[see][]{mihos22};
these data consisted of 14$\times$1200s exposures taken over seven
orbits. The F475W imaging was much shallower, intended only to study the
galaxy's globular cluster system, and had a total exposure time of
2050s, broken into four individual exposures over a single orbit.

\begin{figure*}[]
\centerline{\includegraphics[width=7.5truein]{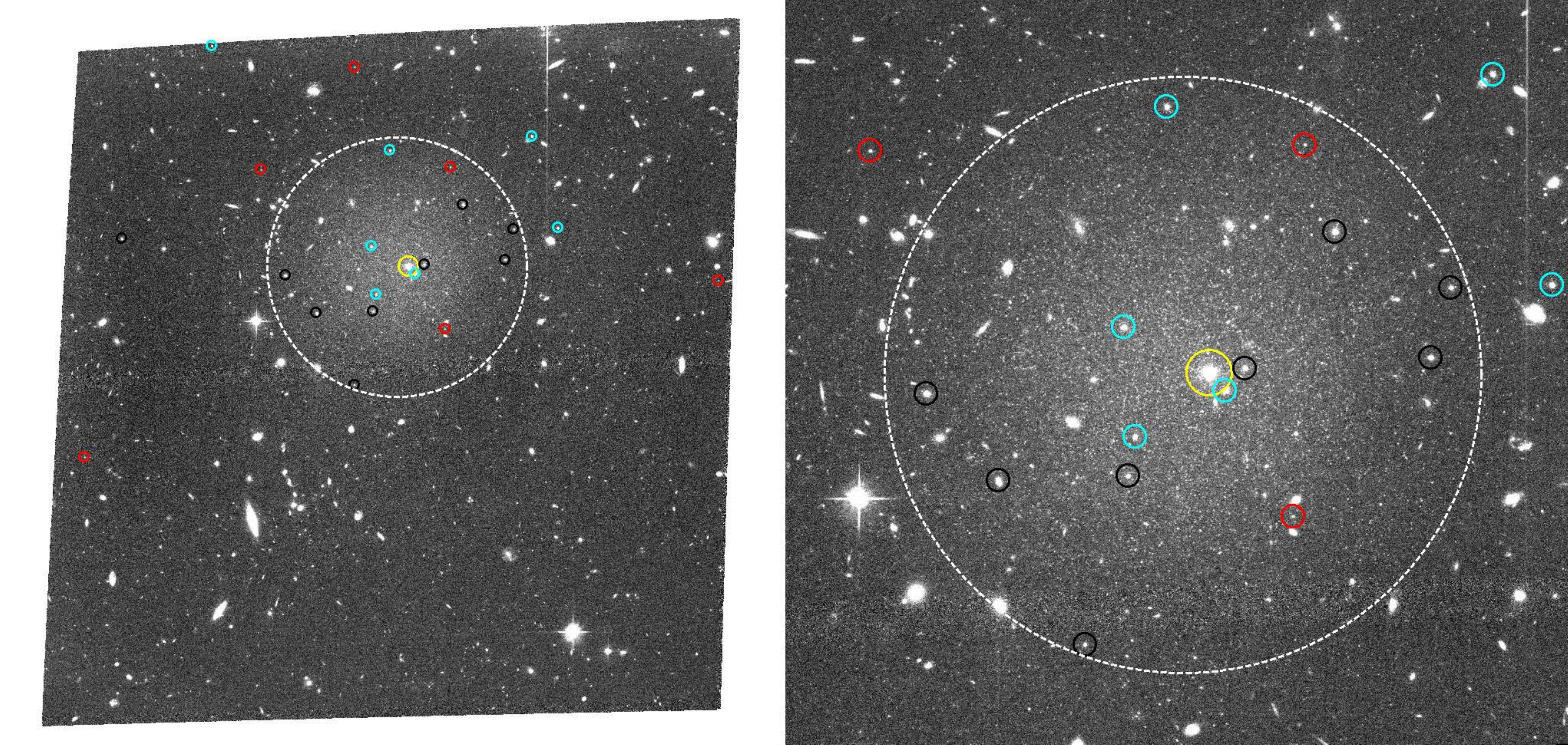}}
\caption{Stacked F814W ACS image of VCC~615, with a total exposure
time of 11$\times$1200s. The left panel shows the full
202\arcsec$\times$202\arcsec ACS field of view, while the right panel is
zoomed in on VCC~615. Our final globular cluster sample is marked in
each panel. Red circles show unresolved candidates ($R_h < 1.3 {\rm
pc}$), black circles show resolved candidates with sizes $1.3 {\rm pc} < R_h < 9.0
{\rm pc}$, and cyan circles show resolved candidates with sizes $R_h > 9 {\rm pc}$.
VCC~615's offset nucleus is shown with a yellow circle, and the large
dotted circle shows a radius of $1.5 R_{e,*}$. The center of the Virgo
Cluster is 1.95\degr\ away (560 kpc in projection) to the east (upper
right) of the field. }
\label{image}
\end{figure*}

For the detection, photometry and characterization of globular cluster
candidates in VCC 615, we created single, deep stacked images in each of
the F475W and F814W filters, using the {\tt astrodrizzle} package within
{\tt drizzlepac}. The details of the creation of the deep, stacked F814W
image were presented in \citep{mihos22} but we will present the salient
points here. Guiding issues during the execution of the program yielded
slightly elliptical point spread functions (PSFs) for most of the
individual, CTE-corrected {\it .flc.fits} images. Of the 14 images
taken, 3 were considered unusable, thus a single, drizzled stacked F814W
image was created from the remaining 11 images, after first
re-registering all of the images using the {\tt tweakreg} and {\tt
tweakback} packages within {\tt drizzlepac} to re-write the WCS for each
image. The resulting image (Figure~\ref{image}) has a combined exposure
time of 13200s, and clearly shows the resolution of the brightest stars
within VCC 615. However, the final PSF of this stacked F814W image is
also slightly elongated, which may slightly reduce our ability to
resolve the smallest cluster candidates in VCC~615.

The shorter exposure times for the individual F475W images resulted in
reduced effects from the trailing issues seen on the F814W images. A
single, deep stacked F475W image was created using the 4 re-registered
CTE-corrected .flc.fits images, with a combined total exposure time of
2050s.

All photometry in this paper has been performed on the stacked images
and has been calibrated to the VEGAmag system using the most recent ACS
zeropoints and corrected for extinction using $A_{F475W} = 0.087$ and
$A_{F814W}=0.041$ \citep{schlafly11}.

\section{Source Photometry and Globular Cluster Candidate Selection}

Before extracting globular cluster candidates from our {\sl Hubble}
imaging, we first assess the depth of the imaging and optimize the
photometric selection of clusters candidates by running a set of
artificial globular cluster tests, using the code \galfit\
\citep{galfit} to inject 100,000 artificial globular clusters into the
imaging. The artificial globular clusters are modeled by a
\citet{king66} profile, sampling evenly in absolute magnitude over the
range $-12.0 \leq M_{\rm F814W} \leq -4.5$, in (log) concentration over
the range $1.0 \leq \log(c) \leq 2.5$, and in (log) physical half light
radius in parsecs over the range $-0.25 \leq \log(R_h) \leq 1.5$. While
this upper bound on size is larger than typical for Galactic globular
clusters, such large clusters have been found in a variety of galactic
environments (see the discussion in Section 5.1.3) and we want to search
parameter space as broadly as possible to identify cluster candidates in
VCC~615. At the distance of VCC~615, these parameters correspond to
apparent magnitudes $19.25 \leq {\rm F814W} \leq 26.75$ and angular half
light radii of $-2.2 \leq \log(r_{\rm e,arcsec}) \leq -0.4$. Before
injection into the imaging, we convolve the model globular clusters with
the image PSF built from a set of bright stars in the ACS field.

Once injected into the images, we use the astropy-affiliated python
package {\tt photutils} \citep{photutils} to map out a local background
on each image and then detect all sources $3\sigma$ above that
background. We look at the size distribution of the recovered sources,
measured using the {\tt photutils} parameter {\tt semimajor\_sigma}. The
clusters are well-bounded by the range $-0.2 \leq \log({\tt
semimajor\_sigma}) \leq 1.0$ pixels, along with a source area of at
least 9 pixels. In terms of selection depth for clusters with sizes $R_h
<$ 10 pc (0.1\arcsec), these tests show that our 50\% completeness limit
is F814W=25.1. For comparison, taking the M87 GCLF peak magnitude of
$\mu_{0,M87}=22.5$ \citep{peng09} and shifting it to the distance of
VCC~615, the expected GCLF peak is at $\mu_{0,VCC615}=22.65$, such that
our imaging should detect clusters 2.5 magnitudes down the luminosity
function. For more extended clusters, the completeness limit is somewhat
brighter, since at fixed magnitude, larger clusters are lower in surface
brightness and thus more difficult to detect. However, even for objects
with $R_h=0.25$\arcsec\ (20 pc), the 50\% completeness limit only rises
to F814W=24.0, still well past the turnover of the GCLF.

Using these artificial globular tests as a guide, we then do the same
automatic source detection procedure on the F475W image to select an
initial sample of globular cluster candidates from the observed data. We
detect all sources $3\sigma$ above the background, and apply the same
cuts on {\tt semimajor\_sigma} and area as described above. We also
require sources to be detected in both the F475W and F814W imaging.
Finally, we add a cut based on source ellipticity. globular clusters are
generally quite round; Galactic clusters all have major:minor axis
ratios of $(b/a)>0.75$ \citep{harris96}, and the vast majority of M87
globular clusters have $(b/a)>0.7$ \citep{madrid09}). Thus our final cut
is to reject elongated sources with axis ratios $(b/a) < 0.65$. After
these cuts, we are left with 298 compact sources detected in both
images, likely consisting of a mix of foreground stars, compact
background galaxies, and globular clusters --- both those specifically
associated with VCC~615, and perhaps others in the general Virgo
intracluster population \citep[\eg][]{durrell14,longobardi18}.

With this initial sample of compact objects in hand, we refine the
sample using a combination of F475W$-$F814W color and F814W aperture
photometry. For each source, we measure aperture magnitudes in both
F475W and F814W using apertures with 3 and 6 pixel radii, applying PSF
aperture corrections \citep{bohlin16} to obtain total magnitudes and
colors for the source. While these aperture magnitudes will
underestimate the total magnitude of more extended sources like globular
clusters, we use them here only for the initial photometric selection of
cluster candidates; when characterizing the properties of the final
cluster sample (Section 5) we use total magnitudes derived from \galfit\
fitting of the sources. For the F814W data, we also calculate the
difference between the 3- and 6-pixel aperture corrected magnitudes,
$\Delta_{(6-3)} \equiv m_6 - m_3$. For unresolved point sources, these
two magnitudes should be equivalent ($\Delta_{(6-3)} \approx 0$), while
for extended sources the additional light at larger radius will yield
$\Delta_{(6-3)} < 0$. Injected clusters in our artificial globular
cluster tests have $\Delta_{(6-3)} \gtrsim -0.8$, and our visual
examination of real sources with $\Delta_{(6-3)} \lesssim -0.8$ show
most objects to be obviously extended background galaxies.

\begin{figure*}[]
\centerline{\includegraphics[width=7.5truein]{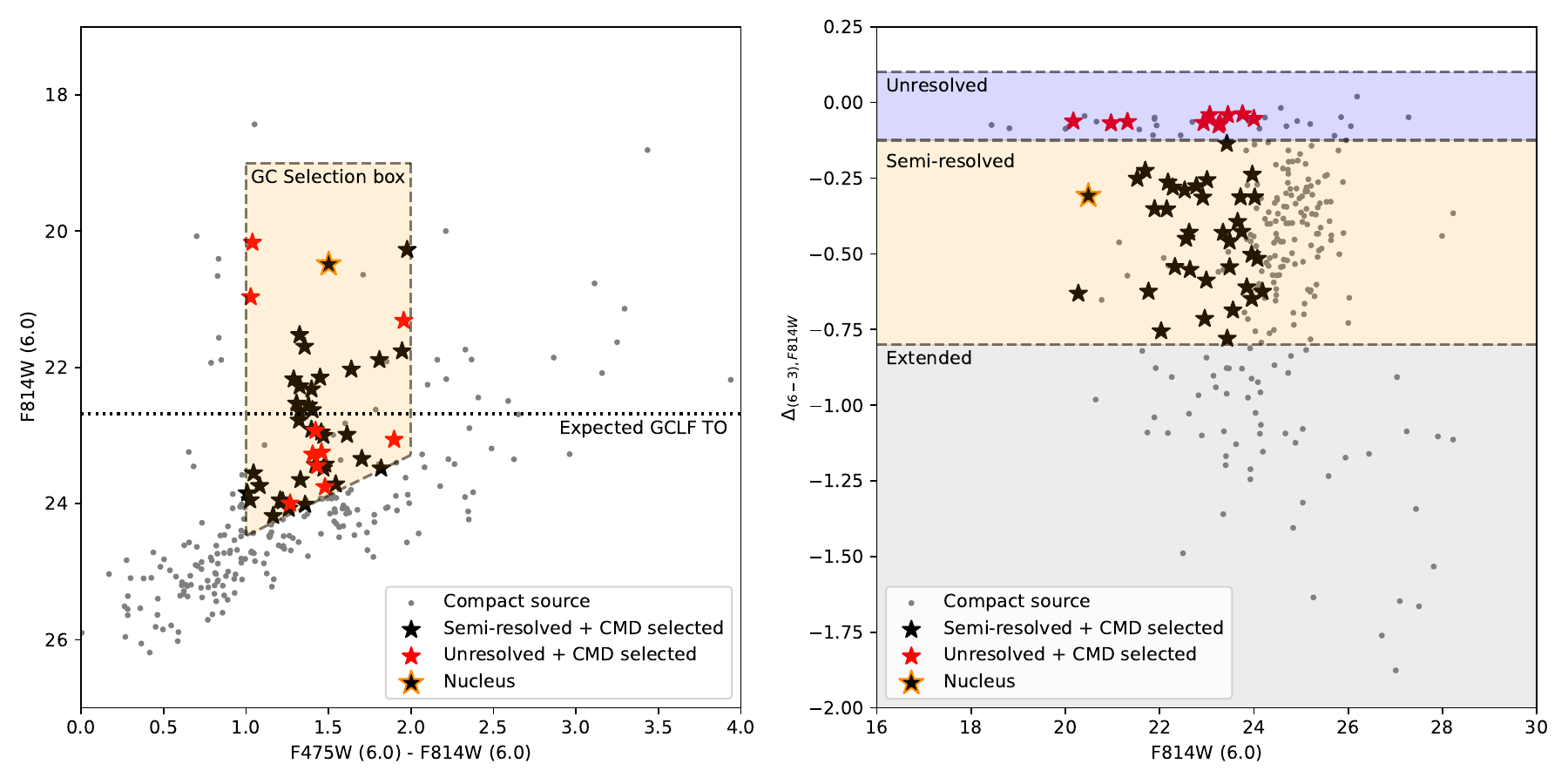}}
\caption{Photometric properties of compact sources. Left panel: F814W vs
F475W$-$F814W color magnitude diagram, with the globular cluster selection box
highlighted. Right panel: $\Delta$F814W (6$-$3) vs F814W, with regions
highlighted for unresolved objects, semi-resolved objects, and extended
objects. Objects which are both color-selected and semi-resolved are
shown as black stars; color-selected unresolved objects are shown as
red stars. The nucleus of VCC~615 is highlighted in orange, and the dotted line
on the CMD shows the M87 GCLF turnover magnitude \citep{peng09}, shifted
to the VCC~615 distance of 17.7 Mpc. 
}
\label{cmd_delta}
\end{figure*}

Using these metrics, we show in Figure~\ref{cmd_delta} the properties of
our initial sample of objects, where the left-hand panel shows the
sources on a color-magnitude diagram (CMD), and the right hand panel
shows $\Delta_{(6-3)}$ vs F814W magnitude. In the CMD, the dashed
horizontal line shows the expected peak magnitude of the VCC~615 GCLF,
obtained by shifting the M87 GCLF peak \citep{peng09} to the distance of
VCC~615. globular clusters are expected to span the color range $1.0 <$
F475W$-$F814W $<2.0$ \citep[see, \eg][]{harris20}, shown in the
highlighted region on the CMD. Within this color range we see a number
of objects, clustered near the peak of the Virgo GCLF. Most of these
objects have relatively blue colors ($1.2 <$ F475W$-$F814W $<1.5$),
consistent with the expectation for metal-poor globular clusters. At
fainter magnitudes (F814W$>23.0$, well below the GCLF turnover), the
selection box becomes increasingly contaminated with background
galaxies, so we place a color-dependent lower limit on the magnitude
selection; at a color of F475W$-$F814W=1.4, typical of metal-poor
globular clusters, this lower limit is F814W=24.0, reaching about 90\%
of the way down the GCLF.

In the distribution of $\Delta_{(6-3)}$ parameter, we see a tight
clumping of objects at $\Delta_{(6-3)} \approx 0$ which represent true
point sources --- foreground Milky Way stars, or other objects compact
enough to be unresolved in our imaging. Below that sequence objects
scatter to more negative $\Delta_{(6-3)}$ values, indicative of more
extended sources. The large number of sources at faint magnitudes
(F814W$>23.0$) and $\Delta_{(6-3)} < -0.25$ again shows the population
of background sources, but many of the brighter sources have
$\Delta_{(6-3)} > -0.8$ where we expect globular clusters to lie. We
thus construct a joint selection for cluster candidates with $1.0 <$
F475W$-$F814W $<2.0$ and $\Delta_{(6-3)} > -0.8$. While this choice will
include unresolved foreground stars in our sample, at this stage of the
selection process we want to avoid excluding any very compact globular
clusters with half light radii $R_h \lesssim$ 1--2 pc. In
Figure~\ref{cmd_delta}, the 46 objects that pass both the color and
compactness criteria are plotted as black stars if they are
semi-resolved ($-0.125 > \Delta_{(6-3)} > -0.8$), and red stars if they
are star-like in compactness ($\Delta_{(6-3)}>-0.125$).

Following the selection of globular cluster candidates based on color
and $\Delta_{(6-3)}$ compactness, we make one final visual check of the
sources. Two of us (JCM and PRD) independently examined the deep F814W
image and found 13 sources to be compact objects embedded in very
diffuse and extended starlight, morphologically similar to the many
background galaxies visible in the field. These types of objects were
found largely at fainter magnitudes, where we expect background
contamination to be high. Furthermore, of the nine objects bright enough
to have deep {\sl NGVS} \ugi\ photometry, all had colors inconsistent
with those of globular clusters (see \S4 below for more details). These
13 sources were thus removed from the candidate list, leaving us with a
total of 33 star-like or slightly resolved globular cluster candidates
to measure half-light radii for using \galfit.

\begin{figure}[]
\centerline{\includegraphics[width=3.5truein]{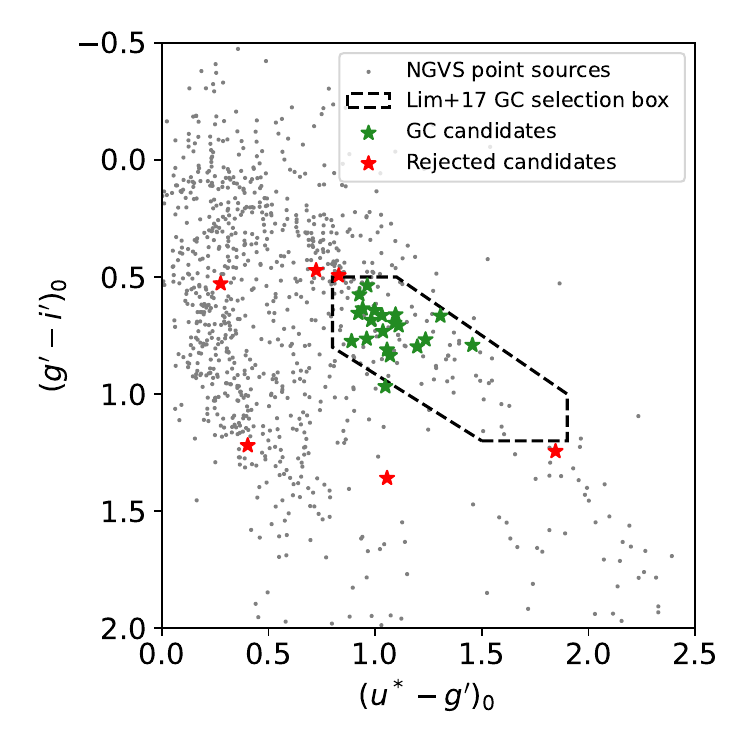}}
\caption{Ground-based NGVS \ugi\  photometry of 
bright (F814W $\lesssim$ 23.5) compact sources
in the VCC~615 field. Small gray dots show all ground-based point sources,
and the dashed polygon is the $u^*, g^\prime, i^\prime$ globular cluster
selection criteria of \citet{lim17}. Our {\sl Hubble}  globular cluster candidates that pass this 
selection are shown as green stars, while those that are rejected are shown as red stars.
}
\label{ugisel}
\end{figure}

The selection of globular cluster candidates thus far has relied
exclusively on the photometry and morphology of the sources in our {\sl
Hubble} imaging. However, for brighter sources we can also take
advantage of additional ground-based photometry in other bandpasses. In
particular, the deep {\sl NGVS} imaging gives us \ugi\ photometry for
sources brighter than F814W $\approx$ 23.5, and the distribution of
sources on the ($u^*-g^\prime, g^\prime - i^\prime$) color plane has
been shown to be an effective selection tool for globular clusters
\citep{munoz14,lim17}. Figure~\ref{ugisel} shows the distribution of our
cluster candidates on the ($u^*-g^\prime, g^\prime - i^\prime$) color
plane, along with all (ground-based) point sources found within a
5\arcmin$\times$5\arcmin\ box around VCC~615 in the {\sl NGVS} imaging.
The dashed region shows the globular cluster selection box defined by
\citet{lim17} based on the colors of the M87 globular cluster
population. Eight objects lie outside the selection box, but one of them
(at $u^*-g^\prime, g^\prime - i^\prime$ $\approx 1,1$) has a background
galaxy projected $<$1\arcsec\ distant which likely contaminates the
ground-based photometry. We retain that source in our candidate list but
reject the other seven, leaving us with a total of 27 cluster candidates
for further analysis.

\section{Sizes of Globular Cluster Candidates}

To measure the half-light radii of our photometrically-selected globular
cluster candidates, we use \galfit\ to model each object using a
S\'ersic model convolved with the PSF derived from bright stars in the
ACS field. We use S\'ersic models rather than a King models because
tests using \galfit\ to fit artificial globular clusters (see below)
showed S\'ersic models to yield a higher frequency of successful fits
than King models, while still accurately measuring the half-light radii
of the clusters. We primarily rely on fitting the F814W imaging, due to
its deeper exposure time and higher signal to noise, but also fit the
F475W imaging as a consistency check on the fitted models. For each
source we make a 250$\times$250 pixel (12.5$\times$12.5 arcsec) image
cutout around the source, masking all objects $3\sigma$ above the
background (save for the source itself). We fit each source with free
parameters being the half-light radius ($R_h$), total F814W magnitude,
S\'ersic-$n$ parameter, and source ellipticity, as well as the locally
estimated background flux level. As initial values, we input $R_h=1$
pixel, $n=4$, a total magnitude equal to the 6-pixel aperture corrected
magnitude, an ellipticity of $b/a=1$ and a sky level set by the overall
median of the fully masked ACS field. The fitting process proved robust
against reasonable variations of these input parameters.

To test the reliability of the \galfit\ estimates of half-light radii
and total magnitude, we used the artificial globular tests and processed
all detected sources through \galfit\ in the manner described above. We
then compared the {\tt galfit}-estimated properties of each source to
its intrinsic (i.e., inserted) properties. For objects with half-light
radii $R_h >$ 0.025\arcsec\ (0.5 pixels), \galfit\ reports a successful
(\ie converged) fit that only slightly overestimates (by about 5\%) the
intrinsic half-light radius. At slightly smaller radii
(0.015\arcsec$<R_h<0.025$\arcsec, or $0.3<R_h<0.5$ pix), \galfit\ does
not report successful convergence, but the fitted sizes remain accurate
to $\pm$20\%). When the intrinsic size is even smaller
($R_h<0.015$\arcsec), the \galfit\ fits fail completely (unconverged
fits with $R_h \approx 0$), signifying an unresolved source. This
0.015\arcsec\ limit on the measurable half-light radius corresponds to a
physical size of 1.3 parsecs at the distance of VCC~615. The estimated
total magnitudes are recovered to within typically $\pm$0.025 mag,
except for objects with extremely small half-light radii
(0.015\arcsec$<R_h<0.025$\arcsec), where uncertainties in the fitted
model lead to the estimated magnitudes being too bright by $\approx$ 0.1
mag.

\begin{figure*}[]
\centerline{\includegraphics[width=7.5truein]{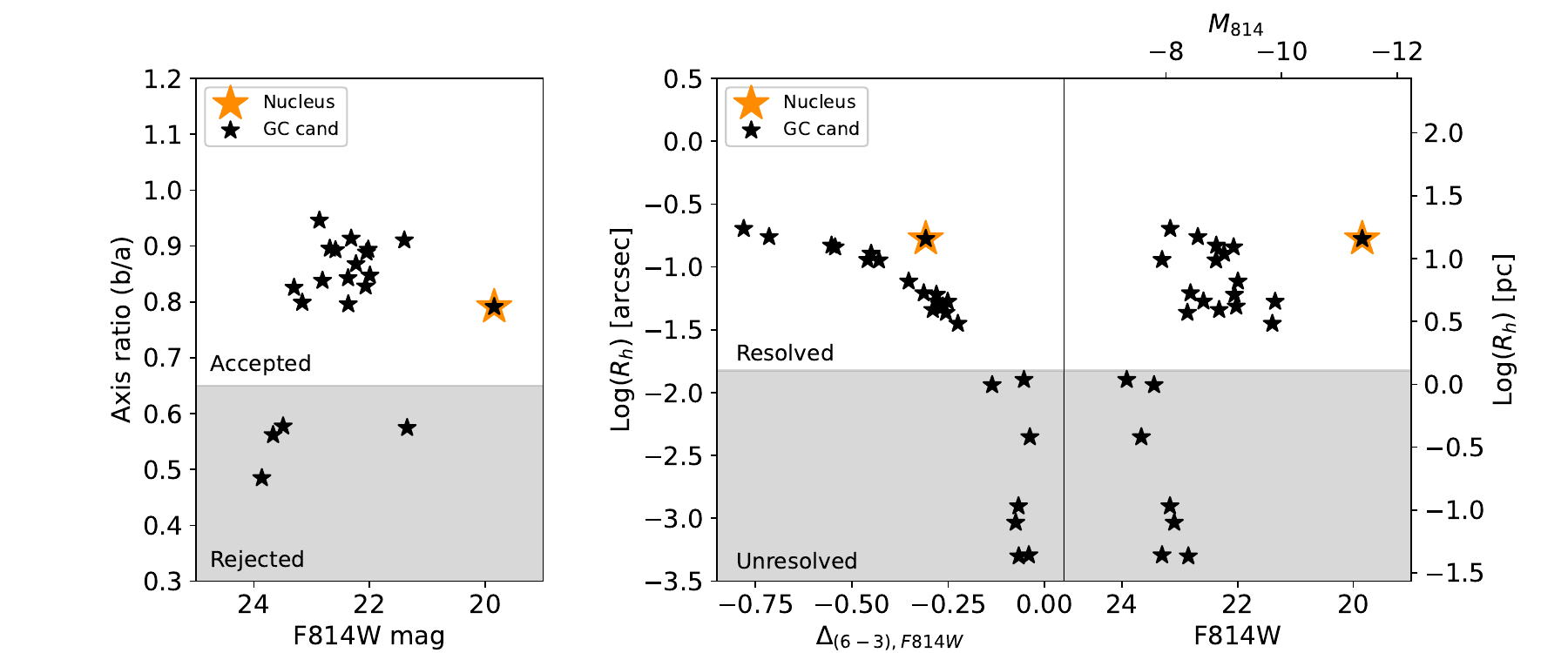}}
\caption{Results of \galfit\ modeling of photometric globular cluster
candidates. Left panel: Ellipticity vs F814W magnitude. The grey shaded
area shows the rejection criteria of $(b/a)<0.65$, although the brightest
object in the rejection region has spectroscopic confirmation from
\citet{toloba23} and is kept in the sample. Right panels: Half-light
radius vs $\Delta_{(6-3)}$ and F814W magnitude for objects which pass the
$(b/a)<0.7$ criterion. In these panels, we plot all the \galfit-reported
half-light radii, but for objects smaller than $R_h=0.015$\arcsec\ (grey
shaded region), these fits are not meaningful and the objects are
considered unresolved. In each panel, the nucleus is highlighted as a
large gold star.
}
\label{galfit}
\end{figure*}

The results of the \galfit\ modeling of the photometric cluster
candidates are shown in Figure~\ref{galfit}. The left panel shows the
fitted ellipticity as a function of F814W magnitude, where we only plot
objects in which \galfit\ reports a successful (i.e., resolved) fit. The
candidates populate two distinct regions: the majority have axis ratios
$b/a \approx$ 0.8-0.95, while a handful show axis ratios $b/a \lesssim$
0.65. We use this ellipticity as an addition selection criterion. Milky
Way globular clusters all have $b/a>0.75$ \citep{harris96}, and
\citet{madrid09} show that the ellipticity distribution of M87 globular
clusters, measured in a very similar manner to what is done here, show a
peak at $b/a=0.9$, with very few flattened clusters at $b/a < 0.6$.
Interestingly, however, the brightest of these flattened sources in our
sample has been confirmed spectroscopically by \citet{toloba23} to have
a Virgo-like velocity of 2090 \kms. We keep this object but reject the
other three, leaving us with 24 globular cluster candidates in our
sample.

The right two panels in Figure~\ref{galfit} show the fitted half-light
radius of each candidate as a function of photometric compactness
($\Delta_{(6-3)}$) and F814W magnitude. In these panels, we plot the
\galfit-reported size even for objects where the fit failed and the
objects are considered unresolved (the gray shaded area in the right two
panels). Two objects fall just below this cutoff at $R_h=0.015$\arcsec
(1.2 pc); while we classify these objects as unresolved, in our
artificial globular tests a small fraction of objects with sizes below
that limit yielded accurate but unconverged fits, and an examination of
the model residuals for these two sources show a clean subtraction. Thus
these objects may be just barely resolved, but for consistency in
building the final sample we place them in the unresolved category.

The other object which does not have a converged fit is the nucleus,
which lies above the otherwise regular sequence of objects where the
fitted size decreases smoothly with the $\Delta_{(6-3)}$ compactness
parameter. The structure of the nucleus is rather complex (see the more
complete discussion in Section~6, following), and at the depth of the
F814W image we start to resolve individual bright stars in the
outskirts, all of which likely complicate the optimization of the fit
parameters. However, the fit residual again shows reasonably clean model
subtraction, and the fits in F814W and F475W --- while both unconverged
--- show good agreement.

We ran two additional tests to assess the robustness of the size
estimates for the globular cluster candidates. We ran \galfit\ on both
the F475W and F814W images, and also independently measured sizes of the
artificial globular clusters on the F814W image using the source fitting
algorithm {\tt ISHAPE} \citep{larsen99}. The lower signal-to-noise in
the F475W image leads to larger uncertainties (and more unconverged
fits) than in the F814W imaging, but both \galfit\ measures follow the
1:1 line with a scatter of only 0.08 dex for objects with good fits.
Comparing the $R_h$ \galfit\ F814W results to those derived from ISHAPE,
we see an even tighter relationship with scatter of only 0.05 dex around
the one to one line for well-fit objects, albeit with increased scatter
for the largest objects. Interestingly, for the two barely-resolved
objects the unconverged the \galfit\ and ISHAPE results both show $R_h
\approx$ 0.25 pixels (0.01\arcsec), again suggesting these objects may
be correctly measured. We also note that while the \galfit\ models did
not fully converge while fitting the nucleus, the \galfit\ F475W and
F814W fits agree and give a clean model subtraction despite reporting
unconverged fits. Attempts to fit the nucleus using ISHAPE were
unsuccessful.

\begin{figure*}[]
\centerline{\includegraphics[width=7.5truein]{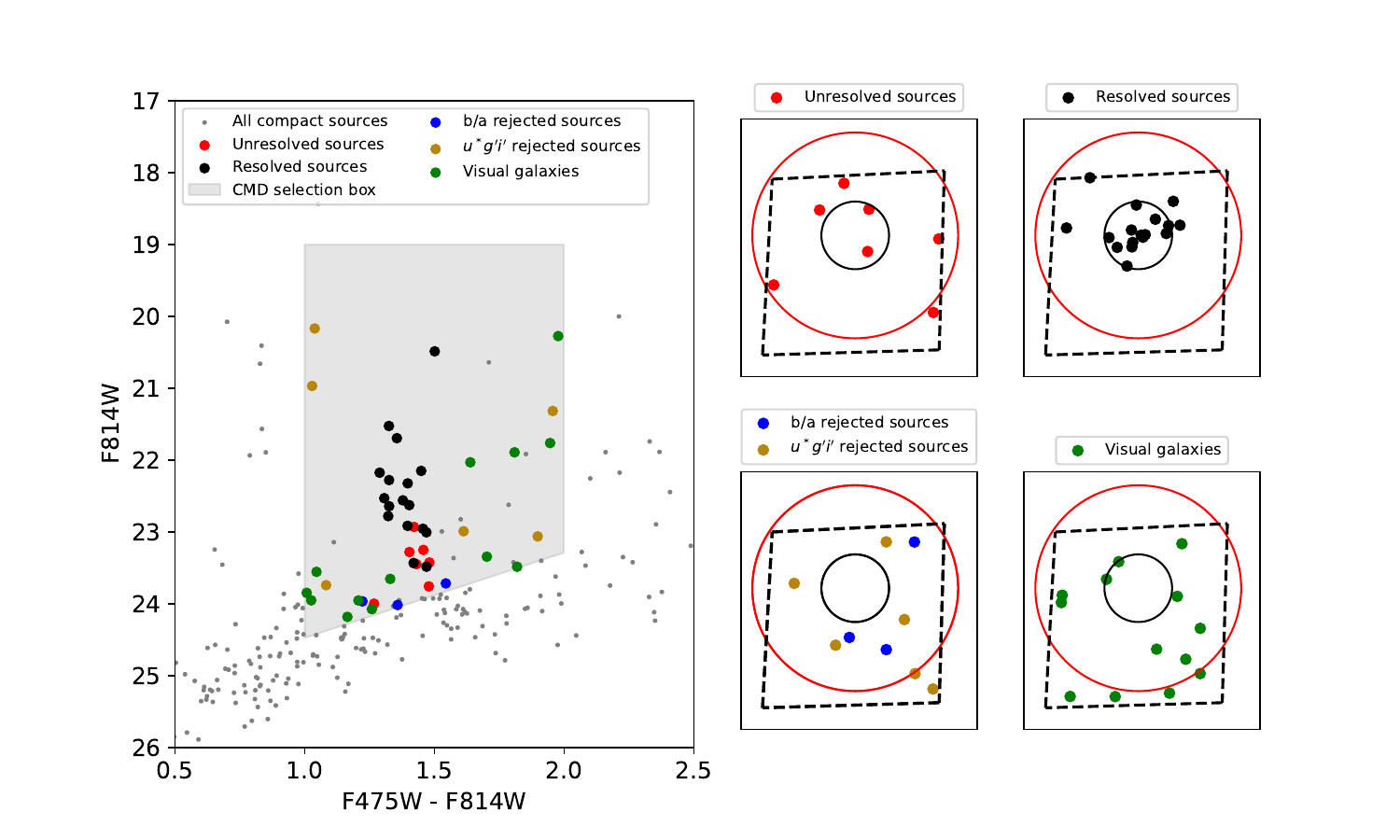}}
\caption{
Photometric and spatial plots of compact sources in the ACS field. Small
grey dots show all compact sources. Larger dots show sources selected
jointly on their location in the CMD and on their $\Delta_{(6-3)}$
compactness factor, color coded by classification. Unresolved sources
(red) have $R_h <$ 0.015\arcsec\ (1.3 pc), while resolved sources
(black) have measured half-light radii in the range $R_h =$ 0.015\arcsec
-- 0.2\arcsec (1.3--17 pc). Objects rejected by their flattened axis
ratio $b/a<0.65$ are shown in blue, objects rejected by their $u^*
g^\prime i^\prime$ colors are shown in magenta, and objects visually
classified as galaxies are shown in green. The inner black circle has a
radius of 1.5$R_{\rm e,*}$ while the outer red circle has a radius of
120\arcsec\ ($\approx 5R_{e,*}$). The field of view of our ACS imaging
is shown by the dashed black box.
}
\label{cmdspat}
\end{figure*}

Armed with size measurements for our globular cluster candidates, we
plot in Figure~\ref{cmdspat} the CMD and sky positions of all the
objects, sorted by classification. The resolved sources (with sizes
$R_h$=0.015--0.2\arcsec, or 1.3--17 pc) cluster tightly both on the CMD
and in the inner 1.5$R_{e,*}$ of the galaxy. The mean color of these
objects (excluding the nucleus) is F475W$-$F814W=1.38 with a scatter of
$\sigma=0.06$ mags. The nucleus itself stands out in the CMD as being
much brighter than the other resolved sources, with a somewhat redder
color of F475W$-$F814W=1.50. Of the two faintest resolved sources, which
start to overlap with the sea of background sources at
F814W$\approx$23.5, one is located within 1.5$R_{e,*}$ and the other (at
the top edge of the ACS field) is found at 3$R_{e,*}$. Both these
objects are kept in the globular cluster sample.

The unresolved sources do not cluster as strongly around the galaxy
spatially, but are mostly found in the same region of the CMD as the
resolved sources. Two of the unresolved sources are found at the
faintest extreme of the CMD selection region, and may be background
sources. One of these is distinct in color from the bulk of the resolved
sources, and is also located just outside the red circle at
$r=$120\arcsec. This object is particularly likely to be a contaminant
and is removed from our globular cluster sample. The other object has a
similar color to the resolved candidates, was one of the two ``just
unresolved'' objects seen in Figure~\ref{galfit}, and is located inside
the $r=$120\arcsec\ circle, at $r\approx 3R_{e,*}$ (the right edge of
the ACS field). The other ``just unresolved'' object is found inside
1.5$R_{e,*}$ at the bottom right. Both these sources, along with the
other four unresolved objects at similar colors in the CMD, are all kept
in our sample of likely globular clusters.

Finally, if we look at the spatial distribution of rejected sources
(visually identified galaxies and compact sources rejected based on
their flattening or $u^* g^\prime i^\prime$ colors), none of those
sources cluster around VCC~615, nor are they similar in their CMD
properties to the resolved and unresolved sources. The flattened sources
are all found at the bottom edge of the CMD selection box, overlapping
with background sources, while the $u^* g^\prime i^\prime$-rejected
sources fall mostly at the extreme edges of the color selection range.
Thus we feel confident that our source rejection criteria are not
rejecting bona-fide globular cluster candidates.

All told, we have a total of 23 objects (17 resolved, 6 unresolved) in
our globular cluster sample. We give the detailed properties of all
objects in our final globular cluster sample in Table~\ref{GCprops}. We
also cross-match the sample against the spectroscopic sample of
\citet{toloba23}, who targeted bright (F814W $\lesssim$ 23) globular
clusters around VCC~615 from the \citet{lim20} sample to derive the
galaxy's dynamical mass. Nine sources (including the nucleus) are in
both our catalog and the \citet{toloba23} spectroscopic catalog. All
nine of those sources lie within $\pm$125 \kms\ of the VCC~615 systemic
velocity \citep[2089 \kms,][]{toloba23}, placing them not only within
the Virgo Cluster, but also almost certainly associating them with
VCC~615 itself. The VCC-like velocities of the spectroscopically
confirmed objects suggests that we have little-to-no contamination from
intracluster globular clusters in our sample.

Finally, we can compare our sample of VCC~615 globular clusters to the
ground-based sample of \citet{lim20}, which uses imaging from the {\sl
Next Generation Virgo Survey} \citep{ferrarese12} to photometrically
select 23 globular cluster candidates within a 120\arcsec\ radius of
VCC~615. Of those 23 sources, three are outside our ACS FOV, two are
resolved as background galaxies, and two are rejected by the \ugi\
selection criteria. The latter two sources are almost certainly
foreground stars: not only are they \ugi-rejected, they also lie at the
extreme color boundaries of the CMD selection region shown in
Figure~\ref{cmd_delta} and are unresolved in our \galfit\ size
estimates. The remaining 16 sources from the Lim cataloged are confirmed
in our catalog as well, including all 12 candidates inside 1.5$R_e$. Our
catalog also contains an additional 7 relatively faint cluster
candidates (F814W$>$23.2) in addition to those in common with the
\citet{lim20} sample.

\begin{figure}[]
\centerline{\includegraphics[width=3.0truein]{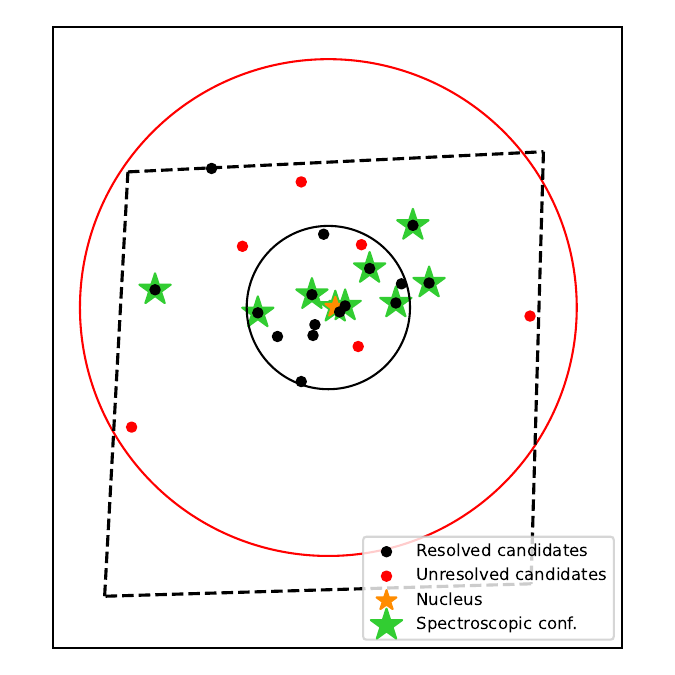}}
\caption{Spatial distribution of our globular cluster candidates in
VCC~615. Black symbols are the semi-resolved candidates, red symbols are
the unresolved candidates. The nucleus is shown with a gold star, while
spectroscopically confirmed sources \citep{toloba23} are shown with a
green star. The black circle has a radius of $1.5R_{e,*}$, the red
circle has a radius of 120\arcsec\ ($\approx 5R_{e,*}$), and the dotted
lines show the ACS field of view.}
\label{spatdist}
\end{figure}

\section{Properties of the VCC~615 globular cluster system}

Here we compare the properties of the VCC~615 globular cluster system to
those in galaxies more generally. Studies of globular clusters in other
UDGs have hinted at differences between their globular cluster systems
and those in more `normal' high surface brightness, but such studies are
often plagued by significant uncertainties, due both to the relatively
small number of globular clusters (and associated high contamination
fractions) and distance uncertainties to the host galaxies. In the case
of VCC~615, our {\sl Hubble} data has provided for a very clean
selection of globular clusters, with a very low contamination fraction.
Furthermore, the galaxy has a well-determined TRGB distance of
$d=17.7^{+0.6}_{-0.4}$ Mpc \citep{mihos22}, significantly reducing the
uncertainties on distance-dependent quantities such as luminosity and
physical size.

In our discussion, we focus first on the intrinsic properties of the
VCC~615 globular clusters themselves, then turn to the connection
between the host galaxy and the globular cluster system as a whole.
Because the nucleus is so distinct photometrically and structurally from
the other cluster candidates (see Section 6, following), in general we
remove it from the sample when discussing the properties of the VCC~615
globular clusters.

\subsection{Properties of the individual globular cluster candidates}

\subsubsection{Luminosity Function}

\begin{figure}[]
\centerline{\includegraphics[width=3.5truein]{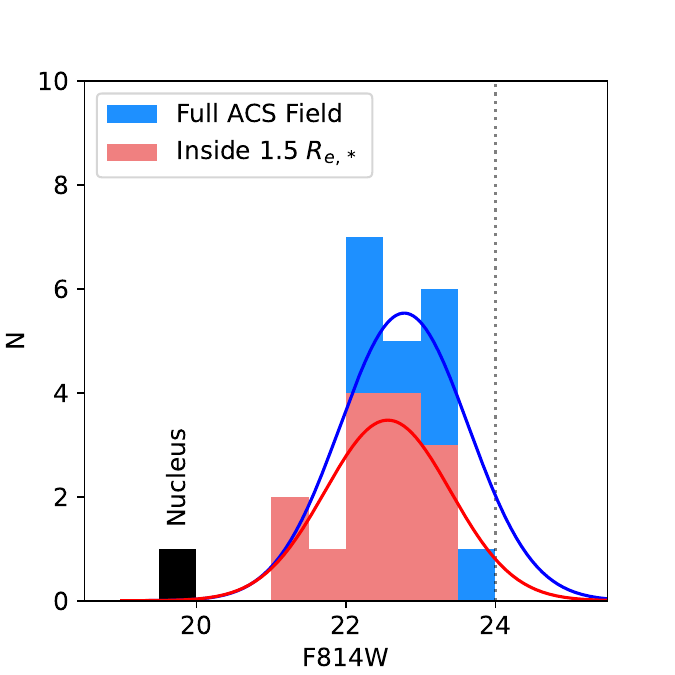}}
\caption{Luminosity function for the globular cluster candidates in
VCC~615, for both the full sample and for objects only within
$1.5R_{e,*}$. The grey dotted line shows the limiting magnitude of the
sample, F814W$_{\rm lim} = 24.0$. The blue and red lines show the
Gaussian fit to the full and inner samples, respectively, with the
nucleus omitted from the fits.}
\label{GCLF}
\end{figure}

We first examine the luminosity function of the globular cluster
candidates, shown in Figure~\ref{GCLF}. Here we break out the sample in
two ways: first by showing all candidates in the ACS field, and also by
showing the subsample of candidates found within $R=1.5R_{e,*}$ of the
photometric center of VCC~615. While the full sample is larger in number
($N_{GC}=22$, not including the nucleus), it is also more likely to
contain a handful of contaminants, predominantly at fainter magnitudes.
Conversely, the inner sample of candidates within $1.5R_{e,*}$ is
smaller ($N_{GC}=14$), but likely has fewer contaminants. When
constructing the luminosity function, we use the \galfit\ F814W
magnitudes for resolved sources successfully fit by \galfit, while for
unresolved objects we use the aperture-corrected six-pixel aperture
magnitudes from {\tt photutils}.

To characterize the luminosity function, we use the Bayesian fitting
package {\tt emcee} \citep{emcee} to model a Gaussian luminosity
function for each sample with turnover magnitude $\mu_0$ and width
$\sigma$ for the two samples, factoring in the magnitude limit of
F814W$_{\rm lim, GC} = 24.0$ set by our selection box in the CMD in
Figure~\ref{cmdspat}. Excluding the nucleus from the fit, we obtain
$\mu_0 = 22.78^{+0.39}_{-0.22}, \sigma = 0.86^{+0.36}_{-0.19}$ for the
full sample, and $\mu_0 = 22.55^{+0.39}_{-0.24}, \sigma =
0.84^{+0.47}_{-0.21}$ for the subsample within $R=1.5R_{e,*}$ (see Table
\ref{GCLFfits}). The difference between the fitted turnover magnitude
between the two samples is within the overall uncertainty of the fit,
and adding the nucleus to the samples does not change the turnover
magnitude in any significant way.

While including the nucleus in the GCLF fit does not change the fitted
turnover magnitude, it does significantly broaden the fitted dispersion,
to values of $\sigma=1.34-1.40$. A GCLF this broad is uncharacteristic
of a low luminosity system like VCC~615. Studies show that the GCLF
width is a declining function of host galaxy luminosity
\citep[\eg][]{durrell96, jordan06, miller07, villegas10}. Dispersions as
large as that seen in our fits that include the VCC~615 nucleus are more
typical of massive ellipticals such as M87 \citep{peng09} and M49
\citep{jordan07} in Virgo, while the lower dispersion for fits without
the nucleus are much more characteristic of low luminosity systems
\citep{lotz04,jordan06, villegas10}. For this reason, as well as the
large gap between the brightness of the nucleus and the next brightest
object in the sample, we adopt the fitted values for $\mu_0$ and
$\sigma$ from the (nucleus-free) full sample as most reflective of the
overall GCLF in VCC~615. Integrating this fitted GCLF over all
magnitudes, we find that our cluster sample is close to fully complete:
only $\approx$ 8\% of the total GCLF lies below our F814W$_{\rm lim, GC}
= 24.0$ magnitude limit.

Studies of globular cluster systems in UDGs have revealed that at least
some systems have populations of overly luminous globular clusters,
skewing the peak of the GCLF to brighter magnitudes and complicating the
use of the GCLF as a distance indicator for these systems. For the
VCC~615 GCLF at a 17.7 Mpc distance, the peak magnitude lies at $M_{\rm
F814W} = -8.45$ for the full sample, and is slightly brighter ($M_{\rm
F814W} = -8.68$) for the inner sample. While this is comparable to the
GCLF peak magnitude for luminous early-type galaxies \citep[$M_{\rm
F814W}\approx-8.5$, \eg][]{kundu01,peng09}, studies show that the peak
magnitude is fainter for low luminosity galaxies. For example,
\citet{miller07} show that the GCLF in Virgo dEs peaks at $M_I=-8.25$
(adjusted for a Virgo distance of 16.5 Mpc), while the relationship
between GCLF peak magnitude and galaxy luminosity derived by
\citet{villegas10} argues that the VCC~615 GCLF peak should be $\approx$
0.3 mag {\it fainter} than in bright ellipticals. Thus our derived GCLF
peak in VCC~615 appears slightly brighter than expected for a galaxy of
its luminosity, by about 0.3--0.5 magnitudes depending on the cluster
sample used (full vs. inner). However, this is close to the $\sim$ 0.3
mag uncertainty in the measurement, and we see no evidence for a large
population of extremely luminous clusters (overluminous by 1.5-2 mags)
such as those found in the UDGs NGC 1052-DF2 and -DF4,
\citep{vandokkum18,shen21a,shen21b}.

\begin{deluxetable}{lccc}
\tablecaption{Gaussian GCLF Fits\label{GCLFfits}}
\tablehead{\colhead{} & \colhead{} & \colhead{All globular clusters} & \colhead{Inside 1.5$R_{e,*}$}}
\startdata
w/ nucleus & $\mu_0$ & $22.93^{+0.56}_{-0.40}$ & $22.54^{+0.61}_{-0.45}$ \\
                 & $\sigma$ & $  1.34^{+0.39}_{-0.34}$ &  $  1.40^{+0.39}_{-0.40}$ \\
\hline
w/o nucleus & $\mu_0$ & $22.78^{+0.39}_{-0.22}$ & $22.56^{+0.40}_{-0.25}$ \\
                   & $\sigma$ & $  0.86^{+0.37}_{-0.19}$ & $  0.84^{+0.48}_{-0.22}$              
\enddata
\end{deluxetable}

\subsubsection{Colors}

Globular cluster colors provide insights into both the characteristic
metallicity and metallicity spread of the cluster population. The color
distribution of the globular clusters in VCC~615 is fairly tight (see
Figure~\ref{cmdspat}) and unimodal, with no evidence of the separate
blue and red populations seen around more massive galaxies
\citep[\eg][]{kundu01,peng06,harris06,lee08,harris23}. Calculating the
colors from the six-pixel aperture magnitudes, for the full cluster
sample we find a mean color of F475W$-$F814W=1.40, with dispersion of
only $\sigma=0.06$ magnitudes, and these values are unchanged if we
consider only the objects in the inner $1.5R_{e,*}$ of the galaxy.

This color is comparable to the blue (more metal-poor) globular cluster
populations seen around much larger galaxies
\citep[\eg][]{peng06,cho16,harris17,harris20,harris23}, and are also
consistent with globular clusters observed in those UDGs with large
enough (typically $N_{GC} > 10$) populations for such measurements.
Using the most recent ACS zeropoints, our color translates to
(F475W$-$F814W)$_{AB} = 0.88$. This value is consistent with the mean
colors of globular clusters in Coma cluster UDGs measured by
\citet{amorisco18} and \citet{saif22} ((F475W$-$F814W)$_{AB}=0.91$ and
0.95, respectively), as well as the $(g-I)_{AB}=0.91\pm0.05$ color for
clusters in the Coma UDG DF17 \citep{bt16}. These colors are, in turn,
similar to that of the blue globular clusters observed in Coma's
intracluster enviroment \citep[$(g-I)=0.89$;][]{peng11}. While our
colors are indicative of a metal-poor population, the values above tend
to be slightly redder (more metal-rich; assuming comparable ages) than
that of dwarf galaxies of similar luminosity. Using the relations
presented by \citet{harris23} our mean color translates to a metallicity
$[Fe/H]\sim -1.2$ (assuming an age of 12 Gyr), which is a value expected
for somewhat higher luminosity systems than VCC~615. This may be
indicative of the higher total masses observed in UDGs with significant
globular cluster populations; our mean metallicity is broadly consistent
with, albeit at the high end, of galaxies with total total masses $\sim
10^{10-11} M_\odot$. This tendency for (at least some) massive UDGs to
have cluster metallicities higher than expected for dwarf galaxies of
similar $M_V$ has also been seen by \citet{bt16} and \citet{janssens22}.

While the color dispersion ($\sigma = 0.06$) in our globular cluster
candidates is small it is not as extremely monochromatic as found in
some other UDGs \citep{muller21,vandokkum22,janssens22,fielder23}, and
there are examples of larger color dispersions in more luminous UDGs
\citep{lim20, saif22}. We see no sign of clusters redder than
F475W$-$F814W = 1.5, where more metal-rich clusters would reside
\citep{harris23}. While a large population of metal-rich clusters is
unlikely in a low luminosity galaxy like VCC~615
\citep[\eg][]{lotz04,peng06} this also argues that we have no
contamination from metal-rich intracluster globular clusters in the
surrounding Virgo environment \citep[\eg][]{durrell14,longobardi18}.

\subsubsection{Sizes}

\begin{figure*}[]
\centerline{\includegraphics[width=7.0truein]{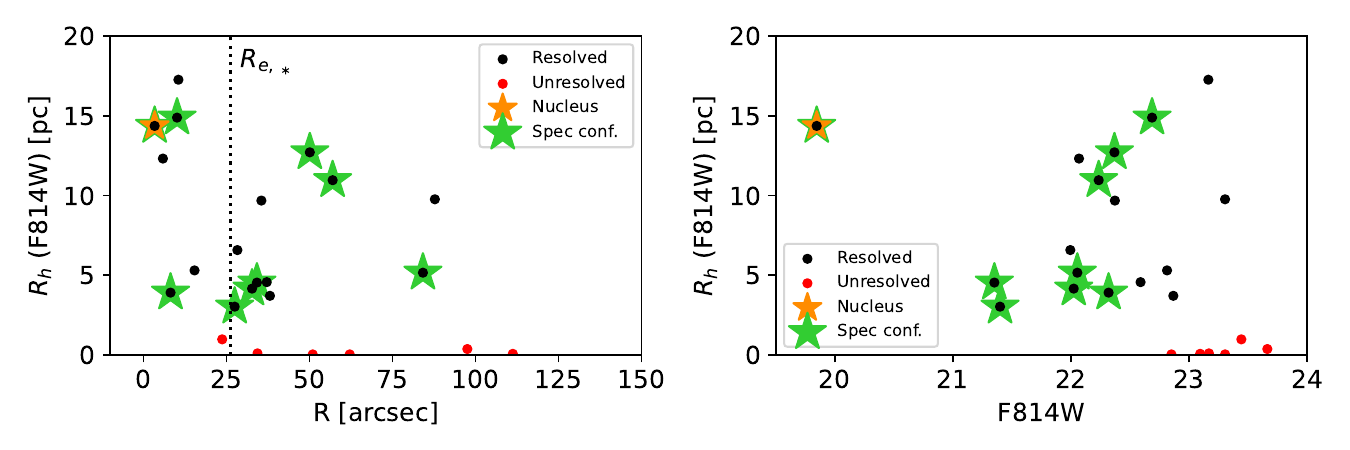}}
\caption{\galfit-measured half-light radii ($R_h$) for the globular
cluster candidates in VCC~615. Left panel: $R_h$ vs galactocentric radius.
Right panel: $R_h$ vs F814W magnitude. Black and red points show resolved
and unresolved sources, respectively, and spectroscopically confirmed
VCC~615 clusters are highlighted with green stars. The nucleus is 
highlighted in gold as well.}
\label{globular clustersizes}
\end{figure*}

While the luminosity function of the VCC~615 globular clusters appear
normal for a galaxy of its luminosity, we do see an excess of overly
large clusters in the sample. Figure~\ref{globular clustersizes} shows
the half-light radii of the globular cluster candidates in VCC~615, as a
function of both galactocentric distance and F814W magnitude. The sizes
show a significant spread, from unresolved sources with $R_h < 1.5$pc,
to larger objects up to $R_h \approx 17$pc. Some of these objects could
in principle be contaminants, but three of them are spectroscopically
confirmed to have Virgo-like velocities within 125 \kms\ of VCC~615
\citep{toloba23}. Furthermore, two other large $R_h$ objects that lack
spectroscopic confirmation are located in the inner 10\arcsec\
($0.4R_{e,*}$) of the galaxy, where contamination is statistically less
likely. Thus it appears probable that many (if not all) of these objects
are large star clusters physically associated with VCC~615.

At first glance the sizes of these large clusters may seem at odds with
those of more typical globular clusters observed in other (larger)
galaxies, where the median half-light radius is $R_h \approx 2.5-3.5$ pc
\citep[\eg][]{larsen01,jordan09,harris09,madrid09,puzia14}. However,
populations of such large ``extended star clusters'' (ESCs) with sizes
$R_h > 7$pc have been found in the outer halos of the Milky Way, M31,
and several other nearby galaxies
\citep{mackey05,huxor05,huxor14,puzia14,mackey19}. These ESCs are
similar in size and luminosity ($M_V>-8$) to the populations of ``faint
fuzzies'' or ``diffuse star clusters'' found in other large galaxies as
well \citep{larsen00,chandar04,peng06,forbes13,liu16}, although those
objects tend to be redder or more metal-rich than the ESCs observed in
the Milky Way or M31. Note here that we distinguish ESCs from objects
like ultra-compact dwarf galaxies (UCDs) and the nuclei of dwarf
galaxies, which are similar in sizes but typically much more luminous.
For example, as shown in Figure~\ref{globular clustersizes}, while the
nucleus of VCC~615 has comparable size to some of the ESCs we find the
galaxy, it is $\approx$ 2.5 magnitudes brighter than the rest of the ESC
population.

In our sample of VCC~615 globular cluster candidates, the fraction of
objects with large half-light radii ($R_h >$ 9 pc) is relatively high:
7/22, or 32\%. This is larger than the fraction of similarly large
clusters found in the Milky Way \citep[30/167 or 18\%, based on the
catalog of][]{baumgardt23}, although the fraction of large clusters
increases in the Milky Way's outer halo \citep[\eg][]{mackey05}. A
similar trend is seen in M31, where $\sim 40\%$ of the clusters at large
projected galactocentric radii ($R_{GC}>25$ kpc) have $R_h > 9$ pc
\citep{mackey19}. Dwarf galaxies also have significant populations of
ESCs of similar size to those we see in VCC~615 \citep[\eg][]{jordan05,
sharina05, georgiev09, dacosta09,hwang11,cole17}; for example, roughly
half the old globular clusters discovered in the M81 group dIrr galaxy
IC 2574 have $R_h>7$ pc \citep{karim24}. Thus the relatively high
fraction of large clusters we find in VCC~615 is similar to that found
in dwarf galaxies and the outskirts of larger galaxies, consistent with
the idea that the weaker tidal fields found in low density environments
are particularly beneficial for the survival of such extended globular
clusters \citep[][]{mackey05,hurley10,hwang11,mackey19}.

Thus the size distribution of globular clusters in UDGs more generally
may also hold information regarding their total mass or, more
specifically, tidal fields. While there are only a few UDGs with
measured globular cluster sizes, there are at least some indications
that they too host significant populations of extended clusters. Studies
of the UDG NGC1052-DF2 have shown that many of its bright,
spectroscopically-confirmed clusters have large half-light radii
\citep{vandokkum18b, trujillo19, ma20}, although these clusters are also
more luminous than the ESCs we find in VCC~615. \citet{muller21} also
find a wide range in $R_h$ for the globular clusters in the UDG
MATLAS-2019. While the brighter, spectroscopically-confirmed clusters in
that galaxy were all reasonably compact ($R_h<7$pc), many fainter
candidates were detected with larger half-light radii, although without
spectroscopic confirmation some of those objects may be background
contaminants. Further studies of globular cluster sizes in UDGs will
thus be extremely important in probing the connection between cluster
properties and their host galaxies, and constraining the dark matter
halos of UDGs.

\subsection{Properties of the globular cluster system}

Turning to the properties of the VCC~615 globular cluster system as a
whole, of particular interest is the total number of globular clusters
$N_{GC}$. Based purely on our observed counts (again, not including the
nucleus), we have 22 globular cluster candidates, including both
resolved and unresolved objects. However, at the faint end of the
distribution, we become susceptible to background contamination; given
the increasing density of background sources in the bottom $\sim$ 0.75
mag of the CMD selection box, we might expect that 2--4 of the faintest
candidates may be contaminants. At the same time, our areal coverage is
only complete out to 75\arcsec\ (3$R_{\rm e,*}$), and we may be missing
a handful of globular clusters that lie outside the ACS field of view
(and indeed, the ground-based study of \citealt{lim20} identified three
cluster candidates that lie just off our ACS field.) Thus, purely from
the raw counts, an observed cluster population of $N_{GC, obs} = 22 \pm
3$ seems reasonable.

We address this question more quantitatively by adopting a model for the
spatial distribution of the globular cluster system. If we use the
Bayesian fitting package {\tt emcee} to fit a S\'ersic model to the
data, including a flat distribution for contaminants: $$\Sigma(R) =
\Sigma_e \exp \Bigl\{ -b_n \Bigl[ \Bigl({R\over
R_e}\Bigr)^{1/n}-1\Bigr]\Bigr\} + \Sigma_b$$ we derive
$n=1.94^{+0.76}_{-0.66}, R_e = 38.8^{+15.2}_{-9.7}$ arcsec,
$\log(\Sigma_e) = -2.3^{+0.74}_{-0.40},
\log(\Sigma_b)=-4.0^{+0.31}_{-0.53}$. The fitted value for S\'ersic-$n$
is consistent with the results for globular cluster systems more
generally, which span the range $1 \lesssim n \lesssim 4$ with a median
value of $n \approx 2$ \citep{lim24}. These fits also provide a total
integrated globular cluster count, including sources outside the ACS
field of view, of $N_{GC,tot} = 23.1^{+6.0}_{-5.0}$ and set the number
of contaminants on the FOV at $N_{cont} = 3.7^{+3.9}_{-2.6}$. These
numbers compare well to the estimates based on the raw counts given
above. Finally, we make one final and small correction for photometric
incompleteness. As shown in the discussion of the GCLF above we have a
completeness fraction of $\approx 0.92$, and factoring this in yields
our final estimate of the total globular cluster population: $N_{GC,tot}
= 25.1^{+6.5}_{-5.4}$. This estimate is slightly smaller than the
\citet{lim20} estimate of $N_{GC,tot} = 30.3\pm9.6$ based on
ground-based imaging, but consistent within the uncertainties.

Previous studies have shown that the globular cluster populations around
UDGs can be quite diverse, likely reflecting the variety of formation
channels that lead to the formation of a UDG. A number of UDGs host
anomalously large numbers of globular clusters for their total
luminosity \citep[\eg][]{lim18, lim20, muller21, danieli22,
fielder23,marleau24}, with specific frequencies as high as $S_N \sim
100-200$. In Virgo, the globular cluster specific frequency in UDGs is
higher than that of the classical Virgo dwarf population \citep{lim20},
and both UDGs and normal dwarfs show increasing $S_N$ values at lower
luminosities \citep{miller07,lim20}. Coupling our VCC~615 globular
cluster count with the total V-band luminosity of the galaxy
($M_V=-14.1$, \citealt{lim20}, adjusted for a 17.7 Mpc distance) we
derive a globular cluster specific frequency of $S_{\rm N} =
55.5^{+14.5}_{-12.0}$, significantly higher than normal Virgo dwarfs,
and even higher than the average for Virgo UDGs of similar luminosity
\citep{lim20}. Following the correlation between globular cluster counts
and total dynamical mass in galaxies \citep[\eg][]{blakeslee97,peng08,
harris13, hudson14, harris17}, such a high specific frequency for
VCC~615 argues for a large mass-to-light ratio for the galaxy,
consistent with that inferred for the galaxy from the kinematics of its
bright globular clusters \citep{toloba18,toloba23}.
 
Indeed we can use VCC~615's total globular cluster count to provide an
independent estimate of the galaxy's total mass. Using the
$N_{GC}$--$M_{tot}$ relationship from \citet{harris17}, VCC~615's total
globular cluster population of $N_{GC,tot}=25.1$ yields a total mass of
$M_{tot}=1.6 \times 10^{11}$ \Msun. Alternatively, adopting a globular
cluster mean mass of $10^5$ \Msun\ \citep{harris17} and using the
\citet{harris17} relationship between total galaxy mass and total
cluster mass, we derive a total mass of $M_{tot}=8.7 \times 10^{10}$
\Msun. Given the galaxy's stellar mass of $M_* = 7.3\pm1.1 \times
10^{7}$ \Msun\ \citep{roedigerIP}, these values indicate a stellar mass
-- halo mass ratio of $\log(M_*/M_h) \approx -3.1\pm0.3$, where the
uncertainty comes largely from the scatter in the globular cluster
scaling relationships \citep{harris17}. Models of the stellar mass --
halo mass relationship \citep[\eg][]{behroozi10,behroozi13,moster13}
argue for higher values ($\log(M_*/M_h) \approx -2.0$ to $-2.5$) for
galaxies of comparable mass to VCC~615. Alternatively, we can adopt a
halo mass profile to calculate the inferred mass-to-light ratio inside
the effective radius of the globular cluster system ($R_{e,GC}$) and
compare it to that measured kinematically by \citet{toloba23}
($M_{1/2}/L_{1/2,V}=148^{+181}_{-144} M_\sun/L_\sun$). Adopting an NFW
halo profile \citep{navarro97} and a halo concentration $c=8$
characteristic for halos of this mass \citep{dutton14} we can scale our
total mass to the mass inside the effective radius. If we use the same
$R_{e,GC}$=23.3\arcsec\ value used by \citet{toloba23}, we derive large
mass-to-light ratios of $M_{1/2}/L_{1/2,V} \approx 99 M_\sun/L_\sun$ and
$79 M_\sun/L_\sun$ for our halo masses derived from the
$N_{GC}$--$M_{tot}$ and $M_{GC}$--$M_{tot}$ relations, respectively. If
we use the larger value of $R_{e,GC}$=38.3\arcsec\ derived from our
sample above, these values rise to $M_{1/2}/L_{1/2,V} \approx 135
M_\sun/L_\sun$ and $116 M_\sun/L_\sun$, respectively. Thus the total
mass and high mass-to-light ratio for VCC~615 inferred from the globular
cluster counts is consistent with the galaxy having an overly massive
dark halo, as argued by \citet{toloba23}.

The fitted size of the VCC~615 globular cluster system ($R_{e,GC} =
38.8^{+15.2}_{-9.7}$ arcsec, or $3.3^{+1.3}_{-0.8}$ kpc) can also
provide insight into the origin of the system. Recent work by
\citet{lim24} derives scaling relationships between the size of globular
cluster systems and the stellar half-light radius and stellar mass of
their host galaxies. In this work, \citet{lim24} shows that these
scaling relationships behave differently for the red and blue globular
cluster sub-systems in galaxies, which are thought to trace clusters
formed {\it in-situ} versus ones later accreted, respectively. At fixed
stellar half-light radius, the size of the VCC~615 globular cluster
system is much smaller than expected for blue (accreted) clusters and
more typical of the red ({\it in-situ}) systems, suggesting that the
globular cluster system in VCC~615 was formed along with the galaxy
itself. However, the tightest correlation found by \citet{lim24} is
between the size of the cluster system and the total number of globular
clusters (Fig 14 of \citealt{lim24}), and here the VCC~615 system
presents as quite large, nearly 2.5 times the size predicted by the
$R_{e,GC}$--$N_{GC,tot}$ relationship. This combination --- relatively
normal in $R_{e,GC}/R_{e,*}$ ratio but high in $R_{e,GC}/N_{GC,tot}$
ratio --- may be a sign that the processes that led to the galaxy being
large for its luminosity (\ie ultradiffuse) may also have led to a
globular cluster system large for its total population. Whether these
properties were imprinted early, as the galaxy formed, or are a result
of ongoing evolution in the cluster environment remains unclear.

\section{VCC~615's offset nucleus}

The brightest compact source in VCC~615 is the galaxy's offset nucleus,
projected just 3.4\arcsec\ (290 pc, or $0.13 R_{e,*}$) northeast of the
galaxy's isophotal center. The nucleus is photometrically distinct from
the other cluster candidates in being significantly more luminous and
somewhat redder in color (Figure~\ref{cmdspat}). Surrounding the nucleus
is a diffuse but semi-resolved envelope of starlight which proved
difficult to fit using \galfit; despite many different trials, we were
unable to achieve a fully converged model in either the F814W or F475W
imaging. In both cases, the fits yielded large values for S\'ersic-$n$
(10.3 and 13.5 in F475W and F814W, respectively), much larger than the
typical values for our cluster candidates ($n\approx 1-5$). These larger
values for S\'ersic-$n$ are qualitatively consistent with the presence
of an extended envelope, and attempts to re-fit the nucleus with a more
globular-like S\'ersic-$n=5$ yielded significantly worse residuals. We
infer that the larger $n$ values are necessary to fit this source, even
if the exact value is not well-determined. However, despite this
uncertainty of the \galfit\ modeling, the object's fitted half-light
radius was quite consistent in both bands, with $R_{h,F475W} =
0.16\arcsec\pm0.01\arcsec$ and $R_{h,F814W} =
0.17\arcsec\pm0.01\arcsec$.

The high luminosity ($M_{F814W}=-11.4$), large size ($R_h =$ 14 pc), and
high S\'ersic-$n$ of this source make it unlike our other globular
cluster candidates and more like a bona-fide nucleus. In Virgo, the
nucleation fraction for galaxies of similar stellar mass to VCC~615
\citep[$\log M_* = 7.86$,][]{roedigerIP} is $\sim$ 60\%
\citep{sanchez19}, so it is unsurprising for the galaxy to be nucleated.
However, offset nuclei are more rare: \citet{binggeli00} estimate that
only 20\% of Virgo dE,N galaxies have offset nuclei, with a weak trend
for the offset to be larger for galaxies at lower surface brightness
\citep[see also][]{poulain21,lambert24}. In some cases, these may be
chance projections of a compact source (a UCD or massive globular
cluster) onto the face of the galaxy, but that is unlikely the case
here. Not only is the nucleus projected only 3.4\arcsec\ from the
center, it also has a radial velocity that is only 4~\kms\ different
from VCC~615's systemic velocity measured by \citet{toloba23}.

So what has caused the nucleus to be offset in VCC~615? One possibility
is that tidal encounters in the Virgo environment have perturbed the
galaxy such that the isophotal center itself is misplaced. This, too,
appears unlikely: the galaxy shows no sign of isophotal irregularity in
ground based imaging by the NGVS \citep{ferrarese12} or the Burrell
Schmidt Deep Virgo Survey \citep{mihos17} projects, and the spatial
distribution of galaxy's resolved stellar populations appears quite
round and symmetric to even lower equivalent surface brightness as well
\citep{mihos22}.

A more speculative possibility is that we are seeing the result of
recent mergers of globular clusters to form the nucleus. If UDGs are
embedded in massive dark halos, dynamical friction will be particularly
efficient at driving globular clusters to the center of the galaxy
\citep{lotz01,dutta20, bar22} where they may merge to form luminous
nuclei, similar to scenarios proposed in normal galaxies
\citep[\eg][]{tremaine75, gnedin14, arca14}. Certainly in the case of
VCC~615, both the dynamical mass estimate \citep{toloba23} and the high
globular cluster specific frequency argue for such a massive dark halo.
Under this scenario, the nucleus being offset from the center may just
be a sign that this sinking process has not completely finished, or that
the central potential is shallow enough that the nucleus can oscillate
around the core for an extended period of time \citep{miller92,taga98}.

Regardless of the origin of the nucleus, its overall properties make it
similar to the population of UCDs found within the Virgo cluster. These
UCDs are typically larger and much more luminous than most globular
clusters \citep[\eg][and references therein]{jones06, brodie11,
chiboucas11, liu15, liu20}, and a number of UCDs have been found to be
embedded in faint and diffuse envelopes
\citep[\eg][]{liu15,voggel16,wang23}, qualitatively similar to the
extended light that gives rise to the high S\'ersic-$n$ value we infer
when fitting this source. If this central source is indeed a nucleus, we
may be seeing the first steps of the evolutionary stripping process
outlined by \citet{wang23}, wherein VCC~615 is slowly being stripped by
gravitational tides in the Virgo Cluster, and may ultimately lead to the
total destruction of the main galaxy leaving behind the nucleus as a
Virgo UCD.

\section{Summary}

We have used F475W and F814W imaging from the ACS camera on the {\sl
Hubble Space Telescope} to study the globular cluster system of the
Virgo Cluster ultradiffuse galaxy VCC~615. Our {\sl Hubble} imaging lets
us construct a very clean and deep sample of globular cluster candidates
that extends more than 90\% down the globular cluster luminosity
function, while simultaneously rejecting background galaxies that would
be unresolved contaminants in ground-based imaging. We follow up the
globular cluster selection by measuring the half-light radii of the
cluster candidates using the image analysis package \galfit\
\citep{galfit}. Using the photometry and size measurements of the
cluster candidates, along with a well-determined TRGB distance for the
galaxy, \citep{mihos22}, we are able to accurately characterize the
physical properties of both the galaxy's globular cluster system and its
offset nucleus. We compare the properties of the VCC~615 globular
cluster system to those of other UDGs, as well as to those in galaxies
more broadly. Our most important results are as follows:

\begin{enumerate}

\item{After selecting on a combination of F475W$-$F814W color, source
size as measured by 3- and 6-pixel aperture magnitudes, and (for
brighter sources) ground-based \ugi\ imaging, we identify 23 globular
cluster candidates in the ACS field of view, down to a limiting
magnitude (set by confusion with background sources) of F814W=24.0. Of
these sources, 15 are located within 1.5 half-light radii of the center
of the galaxy, including the galaxy's compact nucleus. This nucleus is
both photometrically and structurally distinct from the other cluster
candidates, and is removed from our sample to be considered separately.
}

\item{Using \galfit\ we are able to measure half-light radii down to
$R_h = 0.015$\arcsec\ (1.3 pc). Of the 22 globular cluster candidates, 6
are unresolved, while the other 16 have well-determined half-light radii
spanning the range 2--17 pc. The resolved sources are almost equally
split between objects with half-light radii in the range 2--8 pc,
similar to many Galactic globular clusters, and those with much larger
half-light radii in the range 9--17 pc. These latter sources comprise
32\% (7/22) of the total sample of globular cluster candidates and seem
more akin to the ``extended star clusters'' (ESCs) found around a number
of nearby galaxies. Of these 7 large cluster candidates, three are
spectroscopically confirmed to be in Virgo \citet{toloba23}, while two
others are located very near the center of the galaxy (at
$R<0.5R_{e,*}$). Thus the likelihood that this population of sources is
contaminants is very small. }

\item{The luminosity function of the globular clusters in VCC~615 is
characterized by a log normal distribution with a peak magnitude of
F814W $=22.78^{+0.39}_{-0.22}$ ($M_{F814W}=-8.45$ at the 17.7 Mpc
distance of VCC~615) and width $\sigma=0.86^{+0.37}_{-0.19}$. The GCLF
width is consistent with that derived more broadly for galaxies similar
in luminosity to VCC~615, and while the GCLF peak magnitude appears
slightly brighter (by $\approx 0.3$ mag) than expected, we see no
evidence for a population of extremely over-luminous globular clusters
as has been reported in some other UDGs.}

\item{The globular clusters have a mean color of F475W$-$F814W = 1.40,
characteristic of metal-poor stellar populations, and we find no
clusters redder than F475W$-$F814W = 1.5. The dispersion in color is
small $\sigma=0.06$ mag, but not as anomalously monochromatic as the
cluster systems found in some other UDG systems.}

\item{We derive a total globular cluster count for VCC~615 of
$N_{GC}=25.1^{+6.5}_{-5.4}$, after accounting for spatial and
photometric incompleteness. This yields a globular cluster specific
frequency of $S_N=55.5^{+14.5}_{-12.0}$ which is quite large compared to
typical dwarf galaxies of VCC~615's luminosity and implies a large
mass-to-light ratio for the galaxy. Using the scaling relationships
between globular cluster count and total galaxy mass from
\citet{harris17}, we derive a total mass for VCC~615 in the range of
$M_{tot}=0.9$~--~$1.6\times 10^{11}$ \Msun, yielding a ratio of stellar
mass to halo mass of $\log(M_*/M_h) \approx -3.1$. This value is lower
than that predicted by models of the stellar mass -- halo mass
relationship \citep[\eg][]{behroozi13,moster13}, and provides
independent confirmation of the galaxy's high mass-to-light ratio as
derived from the kinematics of its globular cluster system
\citep{toloba18,toloba23}. }

\item{VCC~615's offset nucleus appears photometrically and structurally
distinct from the globular cluster population, and has properties more
akin to those of UCDs. The small offset between the nucleus and the
photometric center of the galaxy (3.4\arcsec\ or 0.13$R_{e,*}$) suggests
that the galaxy is not fully in equilibrium, and we may be seeing a
nucleus recently formed from globular cluster mergers but not yet
dynamically settled into the center of the galaxy.}

\end{enumerate}

Our study adds to the growing census of globular cluster systems in
UDGs. In VCC~615, this census points to a UDG embedded in a massive dark
halo which may guard the galaxy against rapid destruction by the tidal
field of the Virgo Cluster. Instead, slow stripping by repeated passages
through the cluster core may gradually whittle down the galaxy's already
diffuse stellar population, leaving behind only its nucleus to live on
in the cluster as a UCD. The globular clusters
themselves show no sign of being dramatically over-luminous or
monochromatic, as has been found in some other UDGs. However, our size
measurements show that VCC~615 hosts a number of very large globular
clusters with half-light radii $R_h>9$ pc, adding to the variety of
peculiar properties found in the globular cluster systems of UDGs. The
frequency that these globular cluster anomolies occur, and their
connection to the formation and evolution of UDGs, remains an open
question.

\section*{acknowledgments}

We thank Jay Anderson with help and suggestions regarding slightly
trailed {\sl Hubble} images. This research is based on observations made
with the NASA/ESA Hubble Space Telescope for program \#GO-15258 and
obtained at the Space Telescope Science Institute (STScI). STScI is
operated by the Association of Universities for Research in Astronomy,
Inc., under NASA contract NAS5-26555. Support for this program was
provided by NASA through grants to J.C.M. and P.R.D. from STScI. E.T. is
thankful for support from HST GO-15417 and NSF AST-2206498. S.L.
acknowledges the support from the Sejong Science Fellowship Program by
the National Research Foundation of Korea (NRF) grant funded by the
Korea government (MSIT) (No. NRF-2021R1C1C2006790). This research made
use of Photutils, an Astropy package for detection and photometry of
astronomical sources \citep{photutils}.

\facility{HST (ACS), CFHT (MegaCam)} The Hubble Space Telescope imaging data 
used in this study were obtained from the Mikulski Archive for Space Telescopes
(MAST) at the Space Telescope Science Institute and can be accessed via 
\dataset[DOI: 10.17909/z91h-8a19]{https://doi.org/10.17909/z91h-8a19}.
\software{
astropy  \citep{astropy1,astropy2,astropy3}, 
emcee \citep{emcee},
galfit \citep{galfit}
numpy \citep{numpy},
matplotlib \citep{matplotlib},
photutils \citep{photutils},
scipy \citep{scipy},
}

\bibliographystyle{aasjournal}

\begin{thebibliography}{}

\bibitem[Astropy Collaboration et al.(2013)]{astropy1} Astropy Collaboration, Robitaille, T.~P., Tollerud, E.~J., et al.\ 2013, \aap, 558, A33. doi:10.1051/0004-6361/201322068
\bibitem[Astropy Collaboration et al.(2018)]{astropy2} Astropy Collaboration, Price-Whelan, A.~M., Sip{\H{o}}cz, B.~M., et al.\ 2018, \aj, 156, 123. doi:10.3847/1538-3881/aabc4f
\bibitem[Astropy Collaboration et al.(2022)]{astropy3} Astropy Collaboration, Price-Whelan, A.~M., Lim, P.~L., et al.\ 2022, \apj, 935, 167. doi:10.3847/1538-4357/ac7c74
\bibitem[Amorisco \& Loeb(2016)]{amorisco16} Amorisco, N.~C. \& Loeb, A.\ 2016, \mnras, 459, L51. doi:10.1093/mnrasl/slw055
\bibitem[Amorisco et al.(2018)]{amorisco18} Amorisco, N.~C., Monachesi, A., Agnello, A., et al.\ 2018, \mnras, 475, 4235. doi:10.1093/mnras/sty116
\bibitem[Arca-Sedda \& Capuzzo-Dolcetta(2014)]{arca14} Arca-Sedda, M. \& Capuzzo-Dolcetta, R.\ 2014, \mnras, 444, 3738. doi:10.1093/mnras/stu1683
\bibitem[Bar et al.(2022)]{bar22} Bar, N., Danieli, S., \& Blum, K.\ 2022, \apjl, 932, L10. doi:10.3847/2041-8213/ac70df
\bibitem[Barbosa et al.(2020)]{barbosa20} Barbosa, C.~E., Zaritsky, D., Donnerstein, R., et al.\ 2020, \apjs, 247, 46. doi:10.3847/1538-4365/ab7660
\bibitem[Baumgardt \& Vasiliev(2021)]{baumgardt21} Baumgardt, H. \& Vasiliev, E.\ 2021, \mnras, 505, 5957. doi:10.1093/mnras/stab1474
\bibitem[Baumgardt et al.(2023)]{baumgardt23} Baumgardt, H., Sollima, A., Hilker, M. et al.\ 2023, Fundamental parameters of Galactic clusters, https://people.smp.uq.edu.au/HolgerBaumgardt/globular/
\bibitem[Beasley \& Trujillo(2016)]{bt16} Beasley, M.~A. \& Trujillo, I.\ 2016, \apj, 830, 23. doi:10.3847/0004-637X/830/1/23
\bibitem[Beasley et al.(2016)]{beasley16} Beasley, M.~A., Romanowsky, A.~J., Pota, V., et al.\ 2016, \apjl, 819, L20. doi:10.3847/2041-8205/819/2/L20
\bibitem[Behroozi et al.(2010)]{behroozi10} Behroozi, P.~S., Conroy, C., \& Wechsler, R.~H.\ 2010, \apj, 717, 379. doi:10.1088/0004-637X/717/1/379
\bibitem[Behroozi et al.(2013)]{behroozi13} Behroozi, P.~S., Wechsler, R.~H., \& Conroy, C.\ 2013, \apj, 770, 57. doi:10.1088/0004-637X/770/1/57
\bibitem[Bekki et al.(2003)]{bekki03} Bekki, K., Couch, W.~J., Drinkwater, M.~J., et al.\ 2003, \mnras, 344, 399. doi:10.1046/j.1365-8711.2003.06916.x
\bibitem[Benavides et al.(2023)]{benavides23} Benavides, J.~A., Sales, L.~V., Abadi, M.~G., et al.\ 2023, \mnras, 522, 1033. doi:10.1093/mnras/stad1053
\bibitem[Binggeli et al.(1987)]{binggeli87} Binggeli, B., Tammann, G.~A., \& Sandage, A.\ 1987, \aj, 94, 251. doi:10.1086/114467
\bibitem[Binggeli et al.(2000)]{binggeli00} Binggeli, B., Barazza, F., \& Jerjen, H.\ 2000, \aap, 359, 447
\bibitem[Blakeslee(1997)]{blakeslee97} Blakeslee, J.~P.\ 1997, \apjl, 481, L59. doi:10.1086/310653
\bibitem[Blakeslee et al.(1997)]{blakeslee97b} Blakeslee, J.~P., Tonry, J.~L., \& Metzger, M.~R.\ 1997, \aj, 114, 482. doi:10.1086/118488
\bibitem[Blakeslee et al.(2009)]{blakeslee09} Blakeslee, J.~P., Jord{\'a}n, A., Mei, S., et al.\ 2009, \apj, 694, 556. doi:10.1088/0004-637X/694/1/556
\bibitem[Bohlin(2016)]{bohlin16} Bohlin, R.~C.\ 2016, \aj, 152, 60. doi:10.3847/0004-6256/152/3/60
\bibitem[Bradley et al.(2023)]{photutils} Bradley, L., Sip{\H{o}}cz, B., Robitaille, T., et al.\ 2023, Zenodo
\bibitem[Brodie et al.(2011)]{brodie11} Brodie, J.~P., Romanowsky, A.~J., Strader, J., et al.\ 2011, \aj, 142, 199. doi:10.1088/0004-6256/142/6/199
\bibitem[Burkert \& Forbes(2020)]{burkert20} Burkert, A. \& Forbes, D.~A.\ 2020, \aj, 159, 56. doi:10.3847/1538-3881/ab5b0e
\bibitem[Cannon et al.(2015)]{cannon15} Cannon, J.~M., Martinkus, C.~P., Leisman, L., et al.\ 2015, \aj, 149, 72. doi:10.1088/0004-6256/149/2/72
\bibitem[Cantiello et al.(2024)]{cantiello24} Cantiello, M., Blakeslee, J.~P., Ferrarese, L., et al.\ 2024, \apj, 966, 145. doi:10.3847/1538-4357/ad3453
\bibitem[Carleton et al.(2019)]{carleton19} Carleton, T., Errani, R., Cooper, M., et al.\ 2019, \mnras, 485, 382. doi:10.1093/mnras/stz383
\bibitem[Chandar et al.(2004)]{chandar04} Chandar, R., Whitmore, B., \& Lee, M.~G.\ 2004, \apj, 611, 220. doi:10.1086/421934
\bibitem[Chiboucas et al.(2011)]{chiboucas11} Chiboucas, K., Tully, R.~B., Marzke, R.~O., et al.\ 2011, \apj, 737, 86. doi:10.1088/0004-637X/737/2/86
\bibitem[Chilingarian et al.(2019)]{chilingarian19} Chilingarian, I.~V., Afanasiev, A.~V., Grishin, K.~A., et al.\ 2019, \apj, 884, 79. doi:10.3847/1538-4357/ab4205
\bibitem[Cho et al.(2016)]{cho16} Cho, H., Blakeslee, J.~P., Chies-Santos, A.~L., et al.\ 2016, \apj, 822, 95. doi:10.3847/0004-637X/822/2/95
\bibitem[Cole et al.(2017)]{cole17} Cole, A.~A., Weisz, D.~R., Skillman, E.~D., et al.\ 2017, \apj, 837, 54. doi:10.3847/1538-4357/aa5df6
\bibitem[Da Costa et al.(2009)]{dacosta09} Da Costa, G.~S., Grebel, E.~K., Jerjen, H., et al.\ 2009, \aj, 137, 4361. doi:10.1088/0004-6256/137/5/4361
\bibitem[Danieli et al.(2019)]{danieli19} Danieli, S., van Dokkum, P., Conroy, C., et al.\ 2019, \apjl, 874, L12. doi:10.3847/2041-8213/ab0e8c
\bibitem[Danieli et al.(2022)]{danieli22} Danieli, S., van Dokkum, P., Trujillo-Gomez, S., et al.\ 2022, \apjl, 927, L28. doi:10.3847/2041-8213/ac590a
\bibitem[Di Cintio et al.(2017)]{dicintio17} Di Cintio, A., Brook, C.~B., Dutton, A.~A., et al.\ 2017, \mnras, 466, L1. doi:10.1093/mnrasl/slw210
\bibitem[Durrell et al.(1996)]{durrell96} Durrell, P.~R., Harris, W.~E., Geisler, D., et al.\ 1996, \aj, 112, 972. doi:10.1086/118071
\bibitem[Durrell et al.(2014)]{durrell14} Durrell, P.~R., C{\^o}t{\'e}, P., Peng, E.~W., et al.\ 2014, \apj, 794, 103. doi:10.1088/0004-637X/794/2/103
\bibitem[Dutta Chowdhury et al.(2020)]{dutta20} Dutta Chowdhury, D., van den Bosch, F.~C., \& van Dokkum, P.\ 2020, \apj, 903, 149. doi:10.3847/1538-4357/abb947
\bibitem[Dutton \& Macci{\`o}(2014)]{dutton14} Dutton, A.~A. \& Macci{\`o}, A.~V.\ 2014, \mnras, 441, 3359. doi:10.1093/mnras/stu742
\bibitem[Ferrarese et al.(2012)]{ferrarese12} Ferrarese, L., C{\^o}t{\'e}, P., Cuillandre, J.-C., et al.\ 2012, \apjs, 200, 4. doi:10.1088/0067-0049/200/1/4
\bibitem[Ferrarese et al.(2020)]{ferrarese20} Ferrarese, L., C{\^o}t{\'e}, P., MacArthur, L.~A., et al.\ 2020, \apj, 890, 128. doi:10.3847/1538-4357/ab339f
\bibitem[Fielder et al.(2023)]{fielder23} Fielder, C.~E., Jones, M.~G., Sand, D.~J., et al.\ 2023, \apjl, 954, L39. doi:10.3847/2041-8213/acf0c3
\bibitem[Forbes et al.(2013)]{forbes13} Forbes, D.~A., Pota, V., Usher, C., et al.\ 2013, \mnras, 435, L6. doi:10.1093/mnrasl/slt078
\bibitem[Forbes et al.(2020)]{forbes20} Forbes, D.~A., Alabi, A., Romanowsky, A.~J., et al.\ 2020, \mnras, 492, 4874. doi:10.1093/mnras/staa180
\bibitem[Forbes et al.(2021)]{forbes21} Forbes, D.~A., Gannon, J.~S., Romanowsky, A.~J., et al.\ 2021, \mnras, 500, 1279. doi:10.1093/mnras/staa3289
\bibitem[Forbes \& Gannon(2024)]{fg24} Forbes, D.~A. \& Gannon, J.\ 2024, \mnras, 528, 608. doi:10.1093/mnras/stad4004
\bibitem[Foreman-Mackey et al.(2019)]{emcee} Foreman-Mackey, D., Farr, W., Sinha, M., et al.\ 2019, The Journal of Open Source Software, 4, 1864. doi:10.21105/joss.01864
\bibitem[Gnedin et al.(2014)]{gnedin14} Gnedin, O.~Y., Ostriker, J.~P., \& Tremaine, S.\ 2014, \apj, 785, 71. doi:10.1088/0004-637X/785/1/71
\bibitem[Georgiev et al.(2009)]{georgiev09} Georgiev, I.~Y., Puzia, T.~H., Hilker, M., et al.\ 2009, \mnras, 392, 879. doi:10.1111/j.1365-2966.2008.14104.x
\bibitem[Georgiev et al.(2010)]{georgiev10} Georgiev, I.~Y., Puzia, T.~H., Goudfrooij, P., et al.\ 2010, \mnras, 406, 1967. doi:10.1111/j.1365-2966.2010.16802.x
\bibitem[Harris \& van den Bergh(1981)]{hvdb81} Harris, W.~E. \& van den Bergh, S.\ 1981, \aj, 86, 1627. doi:10.1086/113047
\bibitem[Harris(1996)]{harris96} Harris, W.~E.\ 1996, \aj, 112, 1487. doi:10.1086/118116
\bibitem[Harris(2009)]{harris09} Harris, W.~E.\ 2009, \apj, 699, 254. doi:10.1088/0004-637X/699/1/254
\bibitem[Harris(2023)]{harris23} Harris, W.~E.\ 2023, \apjs, 265, 9. doi:10.3847/1538-4365/acab5c
\bibitem[Harris et al.(2006)]{harris06} Harris, W.~E., Whitmore, B.~C., Karakla, D., et al.\ 2006, \apj, 636, 90. doi:10.1086/498058
\bibitem[Harris et al.(2013)]{harris13} Harris, W.~E., Harris, G.~L.~H., \& Alessi, M.\ 2013, \apj, 772, 82. doi:10.1088/0004-637X/772/2/82
\bibitem[Harris et al.(2017)]{harris17} Harris, W.~E., Blakeslee, J.~P., \& Harris, G.~L.~H.\ 2017, \apj, 836, 67. doi:10.3847/1538-4357/836/1/67
\bibitem[Harris et al. (2020)]{numpy} Harris, C.R., Millman, K.J., van der Walt, S.J. et al. \ 2020, Nature, 585, 357
\bibitem[Harris et al.(2020)]{harris20} Harris, W.~E., Brown, R.~A., Durrell, P.~R., et al.\ 2020, \apj, 890, 105. doi:10.3847/1538-4357/ab6992
\bibitem[Harris \& van den Bergh(1981)]{harris81} Harris, W.~E. \& van den Bergh, S.\ 1981, \aj, 86, 1627. doi:10.1086/113047
\bibitem[Hudson et al.(2014)]{hudson14} Hudson, M.~J., Harris, G.~L., \& Harris, W.~E.\ 2014, \apjl, 787, L5. doi:10.1088/2041-8205/787/1/L5
\bibitem[Hunter et al. (2007)]{matplotlib} Hunter, J.D., \ 2007, Computing in Science and Engineering, 9:3, 90.
\bibitem[Hurley \& Mackey(2010)]{hurley10} Hurley, J.~R. \& Mackey, A.~D.\ 2010, \mnras, 408, 2353. doi:10.1111/j.1365-2966.2010.17285.x
\bibitem[Huxor et al.(2005)]{huxor05} Huxor, A.~P., Tanvir, N.~R., Irwin, M.~J., et al.\ 2005, \mnras, 360, 1007. doi:10.1111/j.1365-2966.2005.09086.x
\bibitem[Huxor et al.(2014)]{huxor14} Huxor, A.~P., Mackey, A.~D., Ferguson, A.~M.~N., et al.\ 2014, \mnras, 442, 2165. doi:10.1093/mnras/stu771
\bibitem[Hwang et al.(2011)]{hwang11} Hwang, N., Lee, M.~G., Lee, J.~C., et al.\ 2011, \apj, 738, 58. doi:10.1088/0004-637X/738/1/58
\bibitem[Janssens et al.(2019)]{janssens19} Janssens, S.~R., Abraham, R., Brodie, J., et al.\ 2019, \apj, 887, 92. doi:10.3847/1538-4357/ab536c
\bibitem[Janssens et al.(2022)]{janssens22} Janssens, S.~R., Romanowsky, A.~J., Abraham, R., et al.\ 2022, \mnras, 517, 858. doi:10.1093/mnras/stac2717
\bibitem[Jones et al.(2006)]{jones06} Jones, J.~B., Drinkwater, M.~J., Jurek, R., et al.\ 2006, \aj, 131, 312. doi:10.1086/497960
\bibitem[Jord{\'a}n et al.(2005)]{jordan05} Jord{\'a}n, A., C{\^o}t{\'e}, P., Blakeslee, J.~P., et al.\ 2005, \apj, 634, 1002. doi:10.1086/497092
\bibitem[Jord{\'a}n et al.(2006)]{jordan06} Jord{\'a}n, A., McLaughlin, D.~E., C{\^o}t{\'e}, P., et al.\ 2006, \apjl, 651, L25. doi:10.1086/509119
\bibitem[Jord{\'a}n et al.(2007)]{jordan07} Jord{\'a}n, A., McLaughlin, D.~E., C{\^o}t{\'e}, P., et al.\ 2007, \apjs, 171, 101. doi:10.1086/516840
\bibitem[Jord{\'a}n et al.(2009)]{jordan09} Jord{\'a}n, A., Peng, E.~W., Blakeslee, J.~P., et al.\ 2009, \apjs, 180, 54. doi:10.1088/0067-0049/180/1/54
\bibitem[Karim et al.(2024)]{karim24} Karim, N., Collins, M.~L.~M., Forbes, D.~A., et al.\ 2024, \mnras, 530, 4936. doi:10.1093/mnras/stae611
\bibitem[King (1966)]{king66} King, I.~R.\ 1966, \aj, 71, 64. doi:10.1086/109857
\bibitem[Kundu \& Whitmore(2001)]{kundu01} Kundu, A. \& Whitmore, B.~C.\ 2001, \aj, 121, 2950. doi:10.1086/321073
\bibitem[Lambert et al.(2024)]{lambert24} Lambert, M., Khim, D.~J., Zaritsky, D., et al.\ 2024, \aj, 167, 61. doi:10.3847/1538-3881/ad0f25
\bibitem[Larsen(1999)]{larsen99} Larsen, S.~S.\ 1999, \aaps, 139, 393. doi:10.1051/aas:1999509
\bibitem[Larsen \& Brodie(2000)]{larsen00} Larsen, S.~S. \& Brodie, J.~P.\ 2000, \aj, 120, 2938. doi:10.1086/316847
\bibitem[Larsen et al.(2001)]{larsen01} Larsen, S.~S., Brodie, J.~P., Huchra, J.~P., et al.\ 2001, \aj, 121, 2974. doi:10.1086/321081
\bibitem[Lee et al.(2008)]{lee08} Lee, M.~G., Park, H.~S., Kim, E., et al.\ 2008, \apj, 682, 135. doi:10.1086/587469
\bibitem[Leisman et al.(2017)]{leisman17} Leisman, L., Haynes, M.~P., Janowiecki, S., et al.\ 2017, \apj, 842, 133. doi:10.3847/1538-4357/aa7575
\bibitem[Liao et al.(2019)]{liao19} Liao, S., Gao, L., Frenk, C.~S., et al.\ 2019, \mnras, 490, 5182. doi:10.1093/mnras/stz2969
\bibitem[Lim et al.(2017)]{lim17} Lim, S., Peng, E.~W., Duc, P.-A., et al.\ 2017, \apj, 835, 123. doi:10.3847/1538-4357/835/2/123
\bibitem[Lim et al.(2018)]{lim18} Lim, S., Peng, E.~W., C{\^o}t{\'e}, P., et al.\ 2018, \apj, 862, 82. doi:10.3847/1538-4357/aacb81
\bibitem[Lim et al.(2020)]{lim20} Lim, S., C{\^o}t{\'e}, P., Peng, E.~W., et al.\ 2020, \apj, 899, 69. doi:10.3847/1538-4357/aba433
\bibitem[Lim et al.(2024)]{lim24} Lim, S., Peng, E.~W., C{\^o}t{\'e}, P., et al.\ 2024, \apj, 966, 168. doi:10.3847/1538-4357/ad3444
\bibitem[Liu et al.(2015)]{liu15} Liu, C., Peng, E.~W., C{\^o}t{\'e}, P., et al.\ 2015, \apj, 812, 34. doi:10.1088/0004-637X/812/1/34
\bibitem[Liu et al.(2016)]{liu16} Liu, Y., Peng, E.~W., Lim, S., et al.\ 2016, \apj, 830, 99. doi:10.3847/0004-637X/830/2/99
\bibitem[Liu et al.(2020)]{liu20} Liu, C., C{\^o}t{\'e}, P., Peng, E.~W., et al.\ 2020, \apjs, 250, 17. doi:10.3847/1538-4365/abad91
\bibitem[Longobardi et al.(2018)]{longobardi18} Longobardi, A., Peng, E.~W., C{\^o}t{\'e}, P., et al.\ 2018, \apj, 864, 36. doi:10.3847/1538-4357/aad3d2
\bibitem[Lotz et al.(2001)]{lotz01} Lotz, J.~M., Telford, R., Ferguson, H.~C., et al.\ 2001, \apj, 552, 572. doi:10.1086/320545
\bibitem[Lotz et al.(2004)]{lotz04} Lotz, J.~M., Miller, B.~W., \& Ferguson, H.~C.\ 2004, \apj, 613, 262. doi:10.1086/422871
\bibitem[Ma et al.(2020)]{ma20} Ma, J., Wang, S., Wang, S., et al.\ 2020, \mnras, 496, 3741. doi:10.1093/mnras/staa1775
\bibitem[Mackey \& van den Bergh(2005)]{mackey05} Mackey, A.~D. \& van den Bergh, S.\ 2005, \mnras, 360, 631. doi:10.1111/j.1365-2966.2005.09080.x
\bibitem[Mackey et al.(2019)]{mackey19} Mackey, A.~D., Ferguson, A.~M.~N., Huxor, A.~P., et al.\ 2019, \mnras, 484, 1756. doi:10.1093/mnras/stz072
\bibitem[Madrid et al.(2009)]{madrid09} Madrid, J.~P., Harris, W.~E., Blakeslee, J.~P., et al.\ 2009, \apj, 705, 237. doi:10.1088/0004-637X/705/1/237
\bibitem[Marleau et al.(2021)]{marleau21} Marleau, F.~R., Habas, R., Poulain, M., et al.\ 2021, \aap, 654, A105. doi:10.1051/0004-6361/202141432
\bibitem[Marleau et al.(2024)]{marleau24} Marleau, F.~R., Duc, P.-A., Poulain, M., et al.\ 2024, arXiv:2408.03311. doi:10.48550/arXiv.2408.03311
\bibitem[Mei et al.(2007)]{mei07} Mei, S., Blakeslee, J.~P., C{\^o}t{\'e}, P., et al.\ 2007, \apj, 655, 144. doi:10.1086/509598
\bibitem[Mihos et al.(2017)]{mihos17} Mihos, J.~C., Harding, P., Feldmeier, J.~J., et al.\ 2017, \apj, 834, 16. doi:10.3847/1538-4357/834/1/16
\bibitem[Mihos et al.(2018)]{mihos18} Mihos, J.~C., Carr, C.~T., Watkins, A.~E., et al.\ 2018, \apjl, 863, L7. doi:10.3847/2041-8213/aad62e
\bibitem[Mihos et al.(2015)]{mihos15} Mihos, J.~C., Durrell, P.~R., Ferrarese, L., et al.\ 2015, \apjl, 809, L21. doi:10.1088/2041-8205/809/2/L21
\bibitem[Mihos et al.(2022)]{mihos22} Mihos, J.~C., Durrell, P.~R., Toloba, E., et al.\ 2022, \apj, 924, 87. doi:10.3847/1538-4357/ac35d9
\bibitem[Miller \& Lotz(2007)]{miller07} Miller, B.~W. \& Lotz, J.~M.\ 2007, \apj, 670, 1074. doi:10.1086/522323
\bibitem[Miller \& Smith(1992)]{miller92} Miller, R.~H. \& Smith, B.~F.\ 1992, \apj, 393, 508. doi:10.1086/171523
\bibitem[Moore et al.(1996)]{moore96} Moore, B., Katz, N., Lake, G., et al.\ 1996, \nat, 379, 613. doi:10.1038/379613a0
\bibitem[Moster et al.(2013)]{moster13} Moster, B.~P., Naab, T., \& White, S.~D.~M.\ 2013, \mnras, 428, 3121. doi:10.1093/mnras/sts261
\bibitem[M{\"u}ller et al.(2021)]{muller21} M{\"u}ller, O., Durrell, P.~R., Marleau, F.~R., et al.\ 2021, \apj, 923, 9. doi:10.3847/1538-4357/ac2831
\bibitem[Mu{\~n}oz et al.(2014)]{munoz14} Mu{\~n}oz, R.~P., Puzia, T.~H., Lan{\c{c}}on, A., et al.\ 2014, \apjs, 210, 4. doi:10.1088/0067-0049/210/1/4
\bibitem[Navarro et al.(1997)]{navarro97} Navarro, J.~F., Frenk, C.~S., \& White, S.~D.~M.\ 1997, \apj, 490, 493. doi:10.1086/304888
\bibitem[Peng et al.(2006)]{peng06} Peng, E.~W., C{\^o}t{\'e}, P., Jord{\'a}n, A., et al.\ 2006, \apj, 639, 838. doi:10.1086/499485
\bibitem[Peng et al.(2009)]{peng09} Peng, E.~W., Jord{\'a}n, A., Blakeslee, J.~P., et al.\ 2009, \apj, 703, 42. doi:10.1088/0004-637X/703/1/42
\bibitem[Peng et al.(2008)]{peng08} Peng, E.~W., Jord{\'a}n, A., C{\^o}t{\'e}, P., et al.\ 2008, \apj, 681, 197. doi:10.1086/587951
\bibitem[Peng et al.(2010)]{galfit} Peng, C.~Y., Ho, L.~C., Impey, C.~D., et al.\ 2010, \aj, 139, 2097. doi:10.1088/0004-6256/139/6/2097
\bibitem[Peng et al.(2011)]{peng11} Peng, E.~W., Ferguson, H.~C., Goudfrooij, P., et al.\ 2011, \apj, 730, 23. doi:10.1088/0004-637X/730/1/23
\bibitem[Peng \& Lim(2016)]{peng16} Peng, E.~W. \& Lim, S.\ 2016, \apjl, 822, L31. doi:10.3847/2041-8205/822/2/L31
\bibitem[Pfeffer \& Baumgardt(2013)]{pfeffer13} Pfeffer, J. \& Baumgardt, H.\ 2013, \mnras, 433, 1997. doi:10.1093/mnras/stt867
\bibitem[Poulain et al.(2021)]{poulain21} Poulain, M., Marleau, F.~R., Habas, R., et al.\ 2021, \mnras, 506, 5494. doi:10.1093/mnras/stab2092
\bibitem[Prole et al.(2019)]{prole19} Prole, D.~J., Hilker, M., van der Burg, R.~F.~J., et al.\ 2019, \mnras, 484, 4865. doi:10.1093/mnras/stz326
\bibitem[Puzia et al.(2014)]{puzia14} Puzia, T.~H., Paolillo, M., Goudfrooij, P., et al.\ 2014, \apj, 786, 78. doi:10.1088/0004-637X/786/2/78
\bibitem[Roediger et al., in prep]{roedigerIP} Roediger, J. et al., in prep
\bibitem[Romanowsky et al.(2023)]{roman23} Romanowsky, A.~J., Larsen, S.~S., Villaume, A., et al.\ 2023, \mnras, 518, 3164. doi:10.1093/mnras/stac2898
\bibitem[Saifollahi et al.(2021)]{saif21} Saifollahi, T., Trujillo, I., Beasley, M.~A., et al.\ 2021, \mnras, 502, 5921. doi:10.1093/mnras/staa3016
\bibitem[Saifollahi et al.(2022)]{saif22} Saifollahi, T., Zaritsky, D., Trujillo, I., et al.\ 2022, \mnras, 511, 4633. doi:10.1093/mnras/stac328
\bibitem[S{\'a}nchez-Janssen et al.(2019)]{sanchez19} S{\'a}nchez-Janssen, R., C{\^o}t{\'e}, P., Ferrarese, L., et al.\ 2019, \apj, 878, 18. doi:10.3847/1538-4357/aaf4fd
\bibitem[Sandage \& Binggeli(1984)]{sb84} Sandage, A. \& Binggeli, B.\ 1984, \aj, 89, 919. doi:10.1086/113588
\bibitem[Schlafly \& Finkbeiner(2011)]{schlafly11} Schlafly, E.~F. \& Finkbeiner, D.~P.\ 2011, \apj, 737, 103. doi:10.1088/0004-637X/737/2/103
\bibitem[Sharina et al.(2005)]{sharina05} Sharina, M.~E., Puzia, T.~H., \& Makarov, D.~I.\ 2005, \aap, 442, 85. doi:10.1051/0004-6361:20052921
\bibitem[Shen et al.(2021a)]{shen21a} Shen, Z., Danieli, S., van Dokkum, P., et al.\ 2021, \apjl, 914, L12. doi:10.3847/2041-8213/ac0335
\bibitem[Shen et al.(2021b)]{shen21b} Shen, Z., van Dokkum, P., \& Danieli, S.\ 2021, \apj, 909, 179. doi:10.3847/1538-4357/abdd29
\bibitem[Spitler \& Forbes(2009)]{spitler09} Spitler, L.~R. \& Forbes, D.~A.\ 2009, \mnras, 392, L1. doi:10.1111/j.1745-3933.2008.00567.x
\bibitem[Strader et al.(2006)]{strader06} Strader, J., Brodie, J.~P., Spitler, L., et al.\ 2006, \aj, 132, 2333. doi:10.1086/509124
\bibitem[Taga \& Iye(1998)]{taga98} Taga, M. \& Iye, M.\ 1998, \mnras, 299, 111. doi:10.1046/j.1365-8711.1998.01753.x
\bibitem[Toloba et al.(2018)]{toloba18} Toloba, E., Lim, S., Peng, E., et al.\ 2018, \apjl, 856, L31. doi:10.3847/2041-8213/aab603
\bibitem[Toloba et al.(2023)]{toloba23} Toloba, E., Sales, L.~V., Lim, S., et al.\ 2023, \apj, 951, 77. doi:10.3847/1538-4357/acd336
\bibitem[Tremaine et al.(1975)]{tremaine75} Tremaine, S.~D., Ostriker, J.~P., \& Spitzer, L.\ 1975, \apj, 196, 407. doi:10.1086/153422
\bibitem[Trujillo et al.(2019)]{trujillo19} Trujillo, I., Beasley, M.~A., Borlaff, A., et al.\ 2019, \mnras, 486, 1192. doi:10.1093/mnras/stz771
\bibitem[van Dokkum et al.(2019)]{vandokkum19} van Dokkum, P., Danieli, S., Abraham, R., et al.\ 2019, \apjl, 874, L5. doi:10.3847/2041-8213/ab0d92
\bibitem[van Dokkum et al.(2017)]{vandokkum17} van Dokkum, P., Abraham, R., Romanowsky, A.~J., et al.\ 2017, \apjl, 844, L11. doi:10.3847/2041-8213/aa7ca2
\bibitem[van Dokkum et al.(2018a)]{vandokkum18} van Dokkum, P., Danieli, S., Cohen, Y., et al.\ 2018, \nat, 555, 629. doi:10.1038/nature25767
\bibitem[van Dokkum et al.(2018b)]{vandokkum18b} van Dokkum, P., Cohen, Y., Danieli, S., et al.\ 2018, \apjl, 856, L30. doi:10.3847/2041-8213/aab60b
\bibitem[van Dokkum et al.(2016)]{vandokkum16} van Dokkum, P., Abraham, R., Brodie, J., et al.\ 2016, \apjl, 828, L6. doi:10.3847/2041-8205/828/1/L6
\bibitem[van Dokkum et al.(2015)]{vandokkum15} van Dokkum, P.~G., Abraham, R., Merritt, A., et al.\ 2015, \apjl, 798, L45. doi:10.1088/2041-8205/798/2/L45
\bibitem[van Dokkum et al.(2022)]{vandokkum22} van Dokkum, P., Shen, Z., Romanowsky, A.~J., et al.\ 2022, \apjl, 940, L9. doi:10.3847/2041-8213/ac94d6
\bibitem[Villegas et al.(2010)]{villegas10} Villegas, D., Jord{\'a}n, A., Peng, E.~W., et al.\ 2010, \apj, 717, 603. doi:10.1088/0004-637X/717/2/603
\bibitem[Virtanen et al.(2020)]{scipy} Virtanen, P., Gommers, R., Oliphant, T.~E. et al.\ 2020, Nature Methods, 17, 261
\bibitem[Voggel et al.(2016)]{voggel16} Voggel, K., Hilker, M., \& Richtler, T.\ 2016, \aap, 586, A102. doi:10.1051/0004-6361/201527070
\bibitem[Wang et al.(2023)]{wang23} Wang, K., Peng, E.~W., Liu, C., et al.\ 2023, \nat, 623, 296. doi:10.1038/s41586-023-06650-z
\bibitem[Zaritsky(2022)]{zaritsky22} Zaritsky, D.\ 2022, \mnras, 513, 2609. doi:10.1093/mnras/stac1072
\bibitem[Zaritsky et al.(2023)]{zaritsky23} Zaritsky, D., Donnerstein, R., Dey, A., et al.\ 2023, \apjs, 267, 27. doi:10.3847/1538-4365/acdd71

\end{thebibliography}

\vfil\eject

\centerwidetable
\begin{deluxetable}{cccccccccc}
\tablecaption{VCC~615 Globular Cluster Candidates \label{GCprops}}
\tablehead{\colhead{Obj ID} & \colhead{RA} & \colhead{Dec} & \colhead{F814W} & \colhead{F475W$-$F814W} & \colhead{$R_h$} & \colhead{($b/a$)} & \colhead{M$_{F814W}$} & \colhead{$R_h$} & \colhead{Velocity} \\ 
\colhead{ } & \colhead{(J2000)} & \colhead{(J2000)} & \colhead{$\mathrm{}$} & \colhead{$\mathrm{}$} & \colhead{(arcsec)} & \colhead{$\mathrm{}$} & \colhead{$\mathrm{}$} & \colhead{(pc)} & \colhead{(km s$^{-1}$)} }
\startdata
Nucleus & 185.769724 & 12.015569 & 19.85 $\pm$ 0.01 & 1.50 $\pm$ 0.01 & [0.167 $\pm$ 0.003] & [0.79] & $-$11.39 & [14.4] & 2094 $\pm$ 4 \\
GC01 & 185.762994 & 12.007492 & 21.35 $\pm$ 0.02 & 1.32 $\pm$ 0.01 & 0.053 $\pm$ 0.001 & 0.57 & $-$9.89 & 4.5 &  2091 $\pm$ 12\\
GC02 & 185.776776 & 12.016368 & 21.40 $\pm$ 0.01 & 1.36 $\pm$ 0.01 & [0.035 $\pm$ 0.001] & [0.91] & $-$9.84 & [3.0] & 2122 $\pm$ 15 \\
GC03 & 185.761882 & 12.011469 & 21.99 $\pm$ 0.02 & 1.45 $\pm$ 0.01 & 0.077 $\pm$ 0.003 & 0.85 & $-$9.25 & 6.6 & --- \\
GC04 & 185.774964 & 12.021906 & 22.02 $\pm$ 0.02 & 1.29 $\pm$ 0.01 & 0.049 $\pm$ 0.001 & 0.89 & $-$9.22 & 4.2 & 2075 $\pm$ 7 \\
GC05 & 185.757501 & 11.994417 & 22.05 $\pm$ 0.02 & 1.33 $\pm$ 0.02 & 0.060 $\pm$ 0.001 & 0.89 & $-$9.19 & 5.2 & 2099 $\pm$ 12\\
GC06 & 185.769535 & 12.016412 & 22.07 $\pm$ 0.04 & 1.40 $\pm$ 0.02 & 0.144 $\pm$ 0.012 & 0.83 & $-$9.17 & 12.3 & --- \\
GC07 & 185.785027 & 12.017800 & 22.24 $\pm$ 0.03 & 1.38 $\pm$ 0.02 & 0.128 $\pm$ 0.006 & 0.87 & $-$9.00 & 11.0 & 2169 $\pm$ 33 \\
GC08 & 185.770645 & 12.016551 & 22.32 $\pm$ 0.02 & 1.31 $\pm$ 0.02 & 0.046 $\pm$ 0.001 & 0.91 & $-$8.92 & 3.9 & 2119 $\pm$ 20 \\
GC09 & 185.779822 & 12.024019 & 22.37 $\pm$ 0.06 & 1.33 $\pm$ 0.02 & 0.148 $\pm$ 0.014 & 0.80 & $-$8.87 & 12.7 & 2050 $\pm$ 18\\
GC10 & 185.777010 & 12.008676 & 22.37 $\pm$ 0.05 & 1.40 $\pm$ 0.02 & 0.113 $\pm$ 0.010 & 0.84 & $-$8.87 & 9.7 & --- \\
GC11 & 185.777554 & 12.021043 & 22.59 $\pm$ 0.02 & 1.32 $\pm$ 0.02 & 0.053 $\pm$ 0.001 & 0.89 & $-$8.65 & 4.6 & --- \\
GC12 & 185.769296 & 12.012033 & 22.69 $\pm$ 0.08 & 1.46 $\pm$ 0.03 & 0.173 $\pm$ 0.023 & 0.90 & $-$8.55 & 14.9 & 1966 $\pm$ 41\\
GC13 & 185.764798 & 12.015294 & 22.81 $\pm$ 0.03 & 1.40 $\pm$ 0.03 & 0.062 $\pm$ 0.004 & 0.84 & $-$8.43 & 5.3 & --- \\
GC14 & 185.758685 & 12.017551 & 22.87 $\pm$ 0.02 & 1.47 $\pm$ 0.03 & 0.043 $\pm$ 0.001 & 0.95 & $-$8.37 & 3.7 & --- \\
GC15 & 185.769262 & 12.000680 & 22.93 $\pm$ 0.03 & 1.42 $\pm$ 0.03 & [0.001 $\pm$ 0.002] & [0.53] & $-$8.39 & [0.0] & --- \\
GC16 & 185.766155 & 12.014674 & 23.16 $\pm$ 0.13 & 1.42 $\pm$ 0.04 & 0.201 $\pm$ 0.041 & 0.80 & $-$8.08 & 17.3 & --- \\
GC17 & 185.740219 & 12.002410 & 23.25 $\pm$ 0.03 & 1.46 $\pm$ 0.04 & [0.001 $\pm$ 0.002] & [0.56] & $-$8.15 & [0.1] & --- \\
GC18 & 185.778809 & 12.013634 & 23.28 $\pm$ 0.03 & 1.40 $\pm$ 0.04 & [0.001 $\pm$ 0.006] & [0.08] & $-$8.07 & [0.1] & --- \\
GC19 & 185.775583 & 11.991292 & 23.31 $\pm$ 0.10 & 1.47 $\pm$ 0.05 & 0.114 $\pm$ 0.020 & 0.83 & $-$7.93 & 9.8 & --- \\
GC20 & 185.767108 & 12.021113 & 23.42 $\pm$ 0.03 & 1.48 $\pm$ 0.04 & [0.011 $\pm$ 0.002] & [0.09] & $-$7.80 & [1.0] & --- \\
GC21 & 185.781126 & 12.002181 & 23.45 $\pm$ 0.03 & 1.43 $\pm$ 0.05 & [0.001 $\pm$ 0.001] & [0.69] & $-$7.93 & [0.0] & --- \\
GC22 & 185.784051 & 12.037663 & 23.76 $\pm$ 0.03 & 1.48 $\pm$ 0.06 & [0.004 $\pm$ 0.003] & [0.07] & $-$7.58 & [0.4] & --- 
\enddata
\tablecomments{The listed F814W magnitudes are the fitted total magnitudes from \galfit, except for 
unresolved objects ($R_h<0.015$\arcsec) where we list the 6-pixel aperture corrected magnitudes. The
F475W$-$F814W Vega magnitudes for all objects are derived from the 6-pixel aperture magnitudes. The values for half-light
radius ($R_h$) and ellipticity ($b/a$) come from the \galfit\ modeling; values listed in brackets denote 
unconverged or unresolved fits. Absolute F814W magnitudes and physical half-light radii are derived
using a distance to VCC~615 of 17.7 Mpc \citep{mihos22}, while velocities are taken from \citet{toloba23}.
All magnitudes and colors are corrected for Galactic extinction using values of $A_{\rm F475W}=0.062,
A_{\rm F814W}=0.038$ \citep{schlafly11}.
}
\end{deluxetable}

\end{document}